\documentclass{aa}  
\usepackage{graphicx}
\usepackage{tablefootnote}
\usepackage{txfonts}
\usepackage{url}
\usepackage{xcolor}
\usepackage{threeparttable}
\usepackage{caption}
\usepackage{subfigure}
\usepackage{graphicx}
\usepackage{babel,blindtext}

\usepackage{hyperref}
\begin{document}

   \title{On the Role of Interplanetary Shocks in Accelerating MeV Electrons}

   \subtitle{}

   \author{N. Talebpour Sheshvan\inst{1}, N. Dresing\inst{1}, R. Vainio
          \inst{1}, A. Afanasiev\inst{1}
          \and
          D. E. Morosan
          \inst{2}
          }

        \institute{1. Department of Physics and Astronomy, University of Turku, Finland \\
                   2. Department of Physics, University of Helsinki, Finland\\
              \email{nasrin.talebpoursheshvan@utu.fi}}
        
   \date{}

 
  \abstract
   {One of the sources of solar energetic particle (SEP) events is shocks that are driven by fast coronal mass ejections (CMEs). They can accelerate SEPs up to relativistic energies and are attributed to the largest SEP events. New studies suggest that CME-driven shocks can potentially accelerate electrons to MeV energies in the vicinity of the Sun.}
   {We focus on relativistic electrons associated with strong IP shocks between 2007 and 2019 to determine whether the shocks can keep accelerating such electrons up to 1 AU distance. }
   {We have analyzed High Energy Telescope (HET) observations aboard the STEREO spacecraft of potential electron energetic storm particle (ESP) events, characterized by intensity time series that peak at the time of, or close to, the associated CME-driven shock crossing. We present a new filtering method to assess the statistical significance of particle intensity increases and apply it to MeV electron observations in the vicinity of interplanetary shocks. We employed a STEREO in-situ shock list, which contained in total 587 shocks at the two STEREO spacecraft, from which we identified 27 candidate events by visual inspection. }
   {Our method identified nine clear cases, where a significant increase of MeV electrons was found in association with a shock. Typically, the highest statistical significance was observed in the highest of the three HET energy channels (2.8--4.0 MeV). All nine cases were associated with shocks driven by interplanetary CMEs that showed large transit speeds, in excess of 900 km s$^{-1}$. In several cases multiple shocks were observed within one day of the shock related to the electron increase.}
   {Although electron ESP events at MeV energies are found to be rare at 1 AU our filtering method is not designed to identify a potential interplanetary shock contribution from distances closer to the Sun. Observations by Parker Solar Probe or Solar Orbiter, taken during closer approaches to the Sun, will likely provide clarity on interplanetary shock acceleration of electrons.}

   \keywords{Sun: coronal mass ejections (CMEs) – Sun: particle emission - Acceleration of particles  - shock waves – interplanetary medium}

   \maketitle
%


\section{Introduction}
Shock waves propagating in the interplanetary (IP) space are potential accelerators of energetic particles in the heliosphere. Coronal mass ejections (CMEs) propagating through the IP space are the main drivers of IP shocks, in particular during the solar active years.
Gradual solar energetic particle (SEP)  events \citep[see e.g.,][]{Reames1999, Lee2012, Reames2013, Reames2014, Desai2016, Bruno2018} are thought to be accelerated by CME-driven shocks propagating through the corona and IP space \citep{Lario2008}. 
Energetic storm particle (ESP) events are increases of energetic particle intensities associated with the passage of transient interplanetary shocks \citep[e.g.,][]{Bryant1962, Rao1967, Lario2003, Huttunen-Heikinmaa2009}.

The role of IP shocks for proton intensity enhancements has been investigated in a broad energy range from a few keV to tens of MeV \citep[e.g.,][]{Lario2005, Dresing2016, Makela2011, Ameri2022}. The first ESP events were reported by \citet{Bryant1962} and classified by \citet{Sarris1974} in two categories, spike events and classic ESP events. Spike events are characterized by a sudden rise in proton intensities lasting only between 5 and 20 minutes. In contrast, classic ESP events exhibit gradual intensity increases several hours before the shock passage in agreement with the predictions of the classical diffusive shock acceleration (DSA) theory \citep[]{Desai2016}. \citet{Lario2003, Lario2005} suggested also other types of ESP events based on the features of the intensity profile like ESP+spike or step-like events and those with irregular time-intensity profiles not related to the time of shock crossing.

Although it is well established that protons frequently exhibit classic ESP events, the role of shocks in electron acceleration, especially in IP space, is still not clear. Direct observations of electron acceleration at shocks in IP space are rare \citep[e.g.,][]{Simnett2003, Mitchell_2021}.
Previous studies have analyzed in-situ shock crossings at spacecraft situated at 1 AU and found the shock acceleration efficiency to be very low for electrons at near-relativistic energies, i.e., around 100~keV \citep[]{Tsurutani1985, Lario2005, Makela2011, Dresing2016}. The acceleration efficiency is, however, increasing at lower energies, i.e., tens of keV \citep[]{Yang_2019}.

Although IP shocks at 1 AU do not show clear ability to accelerate electrons in-situ at near-relativistic energies, \citet{Dresing_2022} showed that there is a strong correlation between electron peak intensities observed at Earth’s distance and the Mach number of the shock wave close to the Sun, especially at MeV energies. This provides strong evidence for the ability of coronal shocks to accelerate electrons to relativistic energies and motivates revisiting relativistic electron observations during in-situ shock crossings.

In this study, we present observations of MeV electron ESP events observed with the STEREO mission \citep[e.g.,][]{Kaiser2008} by investigating the entire STEREO dataset for relativistic electron enhancements connected with in-situ shock crossings. Our goal is to determine if IP shocks contribute to MeV electron acceleration. In order to account for acceleration regions that could reside somewhat away from the local region sampled by the spacecraft, we  extended the time interval of the relativistic electron observations to several hours around the time of the shock passage. As a result, a set of candidate events is established, which we then examine more closely to determine the effects of IP shocks on MeV electrons close to 1 AU.

The structure of the paper is the following: in \S\ref{sec:method}, we describe the event selection and a data filtering method developed for identifying shock-related electron enhancements, in \S\ref{sec:results} we present the observations and results of our analysis, and in \S\ref{sec:discon} we discuss the results and present the conclusions of our study.

\section{Event selection and filtering method}\label{sec:method}

To find potential electron ESP events, we used the STEREO in-situ shock list \citep{Jian2013}. The list starts at the beginning of the STEREO mission in January 2007 and the latest update is on April 2019. There are no STEREO-B events after September 2014 because of the loss of contact with the spacecraft. In total, there are 587 interplanetary shocks in the list we used, of which 340 were detected at STEREO-A and 247 at STEREO-B.

At each of the listed in-situ shock-crossing times, we analyzed the electron and proton intensities measured by the High Energy Telescope \citep[HET;][]{von-Rosenvinge2008}. HET is one of four instruments in the Solar Energetic Particle subsystem, which is part of the In-situ Measurements of Particles and CME Transients investigation \citep[IMPACT;][]{Luhmann2008} on STEREO. HET measures the highest energy particles including protons in eleven energy channels from 13.6 to >100 MeV and electrons in three energy channels from 0.7 to 4.0 MeV. In the following, we denote the electron channels as e1 (0.7-1.4 MeV), e2 (1.4-2.8 MeV) and e3 (2.8-4.0 MeV).

We first visually scanned the electron observations for intensity enhancements taking place close in time to the shock arrivals at the spacecraft and, thus, potentially caused by local acceleration of electrons by the shocks. Thus, we found 27 electron ESP candidates. Because the in-situ shock crossings are often associated with high proton and ion fluxes, we needed to be sure that the selected events are not the result of ion contamination in the electron channels. For this, we analyzed 1-minute HET data of the whole year of 2013 (see Appendix \ref{app:contamination} and Fig.~\ref{fig:Nina}). Our conclusion is that the HET electron enhancements are real and not due to ion contamination.
We also checked the radio observations of STEREO/SWAVES to exclude the possibility that the electron enhancements are caused by SEP injections at the Sun coincidentally occurring closely before the enhancements. For this, we searched for type III radio bursts, the signature of accelerated electrons impulsively injected into interplanetary space \citep[]{Reid2014}, occurring about 10-30 minutes before the observed electron enhancement of selected events. Apart from one event (6 Nov 2013, see a more detailed analysis in section \ref{sec:results}), none of our candidates turned out to be associated with a coincident solar injection. 

In our simple scanning method it was not always easy to identify electron intensity enhancements potentially associated with local shock acceleration because they are usually mixed with enhancements of the corresponding SEP events \citep[e.g.,][]{Kouloumvakos2015}. For that reason, we introduced a filtering method for the intensities:  
 \begin{equation}
 \label{eqn:1}
     \Delta I(t_{i};N,M) =  I(t_{i}) - \frac{1}{M} \sum_{j=i-M-N}^{i-1-N} I(t_{j}),
 \end{equation}
 where $M$ is the number of points (time bins) constituting the averaging time window and $N$ is the number of points giving the time lag of the averaging window before each time $t_{i}$. Thus, the method uses the running averaging window with variable length and lag. 
 In the analysis, we took $N = 180$ for all the cases, which corresponds to a constant three-hour lag for one-minute data. The parameter $M$ was given values of 15, 30, 60, 120, 240 and 480, which correspond to a temporal window varying from 15 min to 8 hours for one-minute data.   

\begin{figure*}
    \centering
    \subfigure(a){\includegraphics[width=0.318\textwidth]{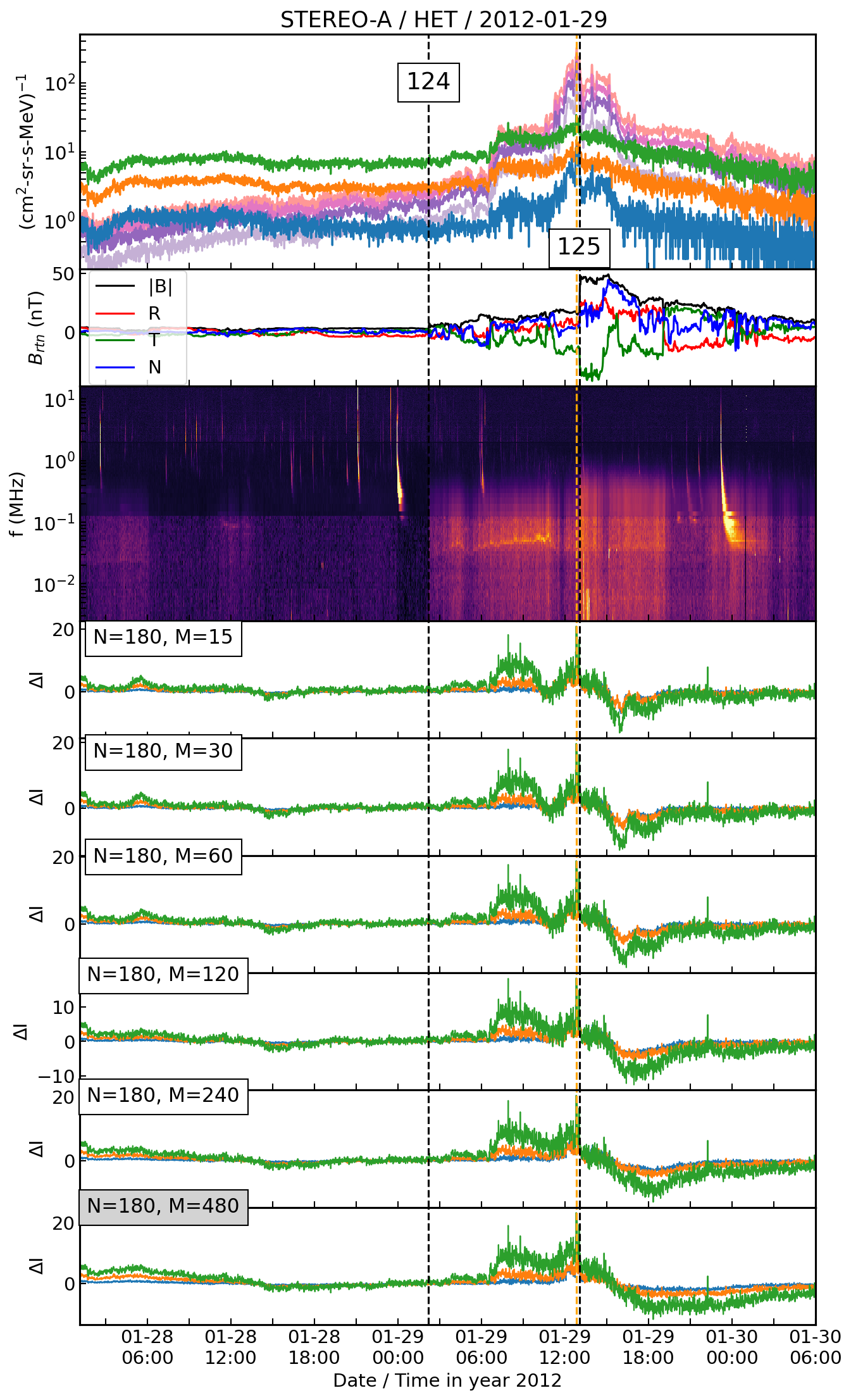}}
    \subfigure(b){\includegraphics[width=0.307\textwidth]{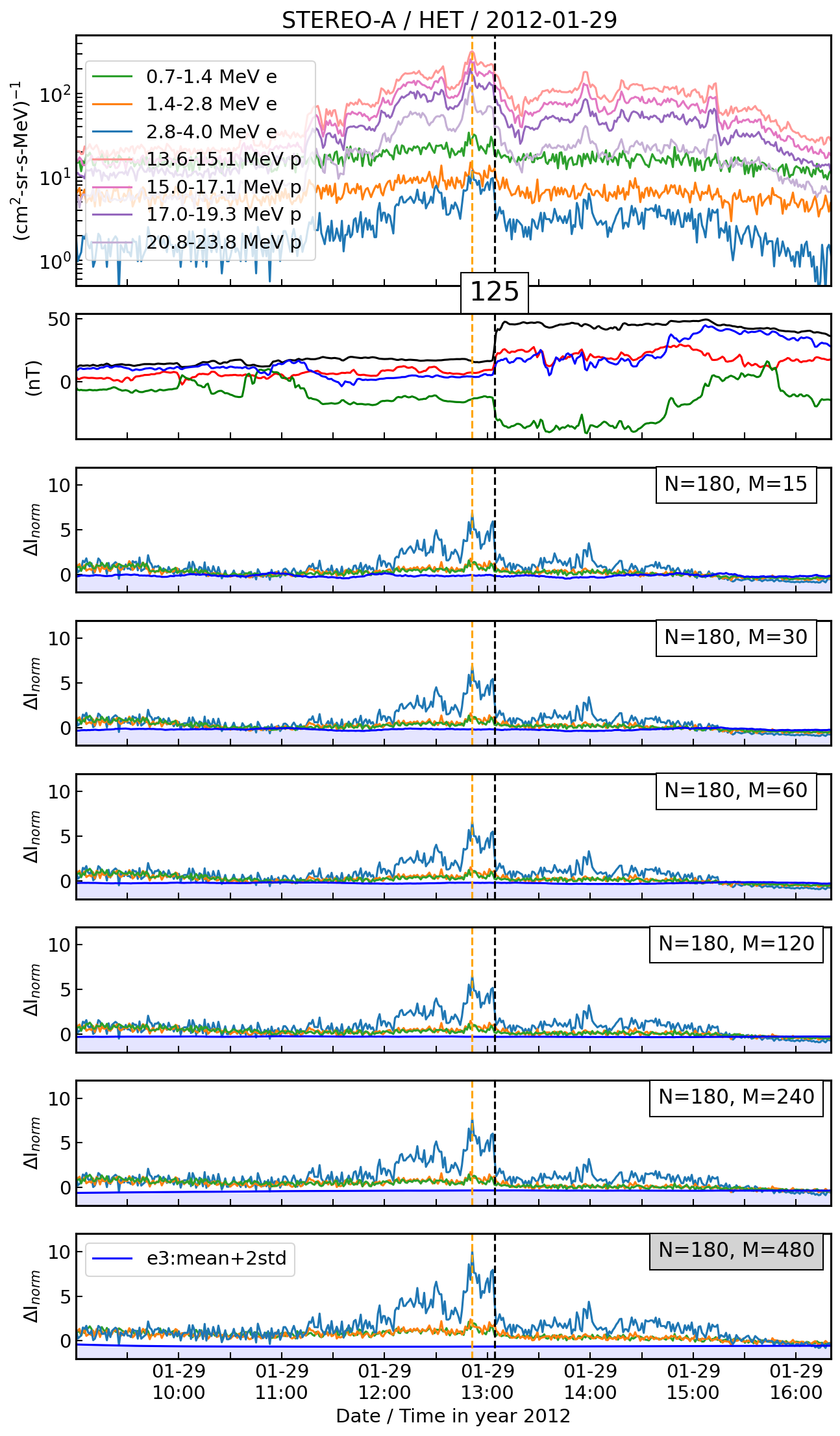}}
    \subfigure(c){\includegraphics[width=0.30\textwidth]{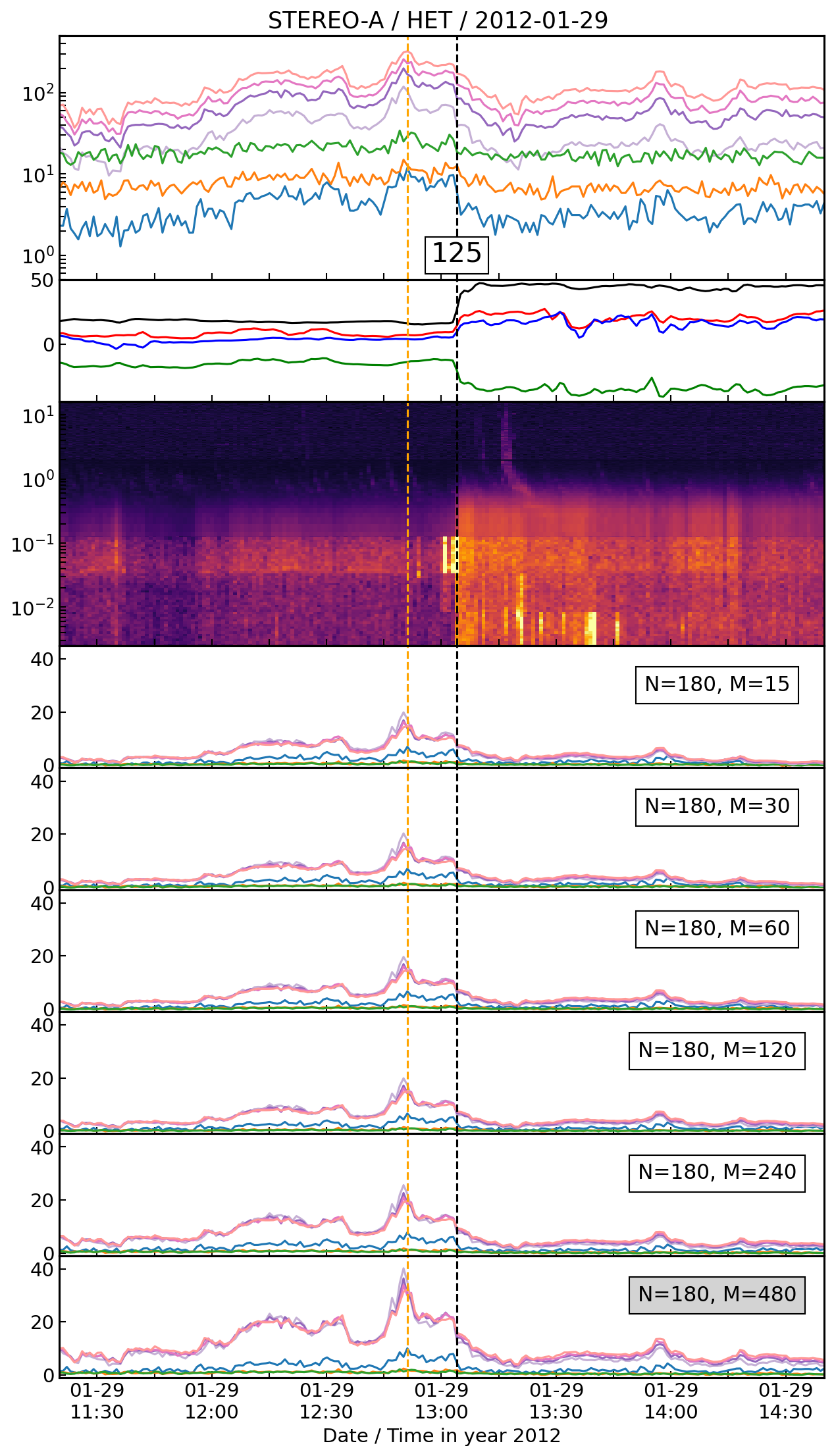}}
    \caption{Analysis of the shock crossing on 29 January 2012 observed by STEREO-A. (a) \textit{From top to bottom}: proton and electron intensity-time profiles as measured by HET, magnetic field vector magnitude and components as measured by IMPACT/MAG, STEREO/WAVES dynamic spectrum, and the rest are filtered data as obtained with the fixed time lag parameter $(N = 180)$ and increasing span parameter $M$. (b) Zoom-in around the time of shock number 125 passage with the electron NF data. (c) Zoom-in around the time of shock number 125 passage with both the electron and proton NF data. The black vertical dashed line indicates the time of the shock passage over the spacecraft, and the orange dashed line marks the time of the peak intensity $I_\mathrm{peak}$ in the electron energy channel having the strongest response in terms of the NF data. Each shaded area in (b) shows the running mean + 2 SD level of the corresponding running background window for the electron energy channel having the strongest response in terms of the NF data.
    }
    \label{fig:125}
\end{figure*}

The applied filtering method allows one to identify more clearly local (in time) variations of the particle intensity. The filtered data are shown in the six bottom panels of Fig.~\ref{fig:125}(a), which also gives an example the analysis carried out for all our 27 electron ESP event candidates. In the intensity time series, one can see intensity enhancements (in various energy channels, for both electrons and protons) associated with the shock and superposed on the ongoing SEP event. The filtering of the intensities conveniently shows the time interval when the intensity variation is strongest.

However, Fig.~\ref{fig:125}(a) also shows that if one wants to compare the filtered intensities corresponding to different energy channels (to find, e.g., which energy channel has the strongest relative enhancement), Eq.~(\ref{eqn:1}) is not convenient due to lower background levels of higher energy channels. Therefore, we modified Eq.~(\ref{eqn:1}) by normalizing the filtered intensity $\Delta I$ by the "local background" intensity, i.e., the intensity value of the running average window:
\begin{equation}
 \label{eqn:2}
    \Delta I_\mathrm{norm}(t_{i};N,M) = \frac{I(t_{i})}{\frac{1}{M}\sum_{j=i-M-N}^{i-1-N} I(t_{j})} - 1. \
\end{equation}
Figure~\ref{fig:125}(b) shows the normalized filtered (NF) intensities $\Delta I_{\rm norm}$ for electrons in the considered example, while Figure~\ref{fig:125}(c) presents both electrons and protons. 
%


In order to determine if an intensity enhancement around the shock-crossing time is significant compared to its preceding intensity level, we calculate the mean and the standard deviation (SD) of the NF data inside the background window, using the six different averaging window lengths (M1-M6, plotted in the lower six panels of the figures) for each of the three electron channels. We then compared the NF data with their corresponding variable background level and calculate the z-score, i.e., the number of SDs above the background level. We applied the z-score to identify statistically significant electron enhancements by requiring that at least three consecutive NF data points within a six-hour window surrounding the shock-crossing time are above the mean + 2 SDs threshold level. From this we identified which energy channel and averaging window length yielded the most significant electron enhancements. 
We then returned to the intensity-time profile of the chosen electron channel and identified the time of its peak intensity $I_\mathrm{peak}$ within a six-hour window surrounding the shock time. Since not all the events have a distinct peak, we divided them into three categories: plateau (Fig.~\ref{fig:126}), peak (Fig.~\ref{fig:141}), and plateau + peak (Fig.~\ref{fig:151}).


Naturally, the NF data also show a strong response at the beginning of the associated SEP events. These intensity enhancements also increase the background levels used in our method, once the background window moves over them. 

In the detailed analyses of example events, we also make use of CME information obtained from the SOHO LASCO CME Catalog \citep[\url{http://cdaw.gsfc.nasa.gov/CME_list}; e.g.,][]{2004JGRA..109.7105Y, 2009EM&P..104..295G, 2010SunGe...5....7G}. 

\section{Observations and Results}\label{sec:results}
%
\begin{table*}[h!]
    \caption{Parameters of the nine electron ESP events. Associated shock parameters from \url{https://stereo-ssc.nascom.nasa.gov/pub/ins_data/impact/level3/IPs.pdf}.}             
        \label{table:1}      
    \centering          
    \begin{tabular}{c c c c c c c  l l l l}     
    \hline\hline       
    Date & Sh num./SC & $N_{sh}$\textsuperscript{1} & $V_{t}$ [km/s]\textsuperscript{2} & $M_{ms}$ & $\theta_{Bn} [^\circ]$ & $r_{B}$\textsuperscript{3} & $e_{r}$\textsuperscript{4} & $\Delta t$ [min]\textsuperscript{5} & shape & $M$\textsuperscript{6}\\ 
     (1) & (2) & (3) & (4) & (5) & (6) & (7) & (8) & (9)  & (10) & (11)\\ 
    \hline 
     2011-06-05 & 87 / A & (1,0)\textsuperscript{*} & 1749&1.54 & 52.60 & 1.57 & e1 & +66& peak & 60 \\
     2012-01-29 & 125 / A & (0,0) & 917& 2.00 & 87.70 & 2.15 & e3 & $-$13 & peak & 480 \\
    2012-03-08 & 126 / B & (1,1)\textsuperscript{*} & 1147&1.48 & 58.70 & 1.60 & e3 & $-$92 & plateau & 480\\
   2012-05-28 & 141 / A & (0,0) & 1321 & 2.85 & 74.80 & 2.70 & e3 & $-$4  & peak  & 480\\
   2012-07-23 & 151 / A & (0,0) & 2099 & 2.46 & 45.50 & 2.17 & e3 & $-$1 &  plateau+peak & 480 \\
   2013-11-06 & 198 / B & (1,1) & 989 & 1.92 & 64.30 & 2.02 & e3 & +31 & peak & 15\\  
   2014-09-25 & 246 / B & (2,1) & 2114 & 1.56 & 55.30 & 1.60 & e1 & +132 & peak & 480\\
   2017-07-24 & 316 / A & (0,2) & 1197 & 2.00 & 47.92 & 1.92 & e3 & 0 & plateau & 240\\
   2017-09-19 & 322 / A & (0,2) & 985 & 2.59 & 44.90 & 2.33 & e3 & $-$8 & peak & 15\\
    \hline                  
    \end{tabular}
\footnotesize
\begin{tablenotes}
\item[Note:] \textsuperscript{1}The first (second) number in each bracket indicates how many shocks were observed within two days before (after) the corresponding shock indicated in column (2).\\  \textsuperscript{2} Transit speed of the associated CME.\\ \textsuperscript{3} Magnetic compression ratio of the shock.\\ \textsuperscript{4} The electron energy channel showing the strongest signal in terms of of NF data. \\ \textsuperscript{5} $\Delta t$ between the peak intensity time and the shock-crossing time. Negative numbers refer to peaks in the upstream region and positive numbers indicate peaks in downstream region of the shock.\\  \textsuperscript{6} Best averaging time window for each event.
\item[*] \textsuperscript{*}SIR shock occurs one day before the shock crossing time.  
\end{tablenotes}
\end{table*}

The filtering technique (described in \S\ref{sec:method}), applied to the visually identified candidate events in STEREO/HET observations, enabled us to list IP shocks potentially accelerating electrons to MeV energies. Out of the 27 candidate IP shock events observed by STEREO A and B, nine ($\sim$1.5\%) turned out to have significant relativistic electron intensity enhancements. 
The first and second column of Table~\ref{table:1} list the event date, shock number \footnote{\url{https://stereo-ssc.nascom.nasa.gov/pub/ins_data/impact/level3/IPs.pdf}}, and the observing spacecraft (A or B), respectively. Two numbers are indicated in parentheses in the third column; the first one gives the number of shocks that occurred in the two days prior to our main shock, and the second number provides the number of shocks that occurred in the two days following our main shock. The shock transit speed calculated using the start time of the corresponding type III radio burst observed by STEREO/WAVES and the CME arrival time \footnote{\url{https://stereo-ssc.nascom.nasa.gov/pub/ins_data/impact/level3/ICMEs.pdf}} is listed in the fourth column.
The shock parameters, including the magnetosonic Mach number, shock normal angle, and shock magnetic compression ratio, given in the fifth, sixth, and seventh columns, respectively, were taken from the STEREO IP shock list\footnote{\url{https://stereo-ssc.nascom.nasa.gov/pub/ins_data/impact/level3/IPs.pdf}}. The last four columns contain information regarding the shock-associated electron enhancements. Column (8) gives the energy channel with the highest z-score in the NF data among the three electron channels. Column (9) gives the time delay between the intensity peak time and the shock crossing time (negative value indicates that the intensity peaks before the shock-crossing time). The shape of the intensity-time profile around the shock crossing (visually evaluated) is given in column (10), and in the last column of the Table~[\ref{table:1}], the background window length parameter $M$ corresponding to the highest z-score in the related electron channel, in the 6 hours window around the time of shock crossing, is presented. 



Applying the normalization to the filtered data allows us to identify the energy channel with the strongest response on the events. In seven of these events the most significant signal belongs to the e3 channel of HET in the range of 2.8-4.0 MeV (Table \ref{table:1}). In six cases ($\sim$67\%) the peak intensity is observed upstream of the shock, with five of them occurring very close to the time to the shock crossing (less than 20 minutes). 
In four cases (shock numbers 87, 126, 151, and 246, see Fig. \ref{fig:151}, \ref{fig:246}), the MeV electron channels show overall higher intensities than the deka-MeV proton channels around the time of the shock crossings. However, in the remaining five cases (shocks number 125, 141, 198, 316 and 322) the overall proton intensities are higher. In two events (shocks number 125 and 246), the signal of the NF data of the proton channels are stronger than the those of the electron channels (Fig. \ref{fig:125} and \ref{fig:246} (c)). However, in the rest of events the response of the MeV electrons, as measured by the NF data, is stronger than that of the protons.
The mean value of the magnetosonic Mach number of the nine analyzed in-situ shocks is 2.04, and is varying from 1.48 to 2.85. These are not outstanding values among IP shocks. The shock normal angles $\theta_{B_{n}}$ vary from 45.5° to 87.7°, i.e., from oblique to quasi-perpendicular. The majority of the shocks are oblique shocks. Only three shocks are quasi-perpendicular with shock normal angles larger than 60°.

\begin{figure*}
   \centering
   \includegraphics[width=0.75\textwidth]{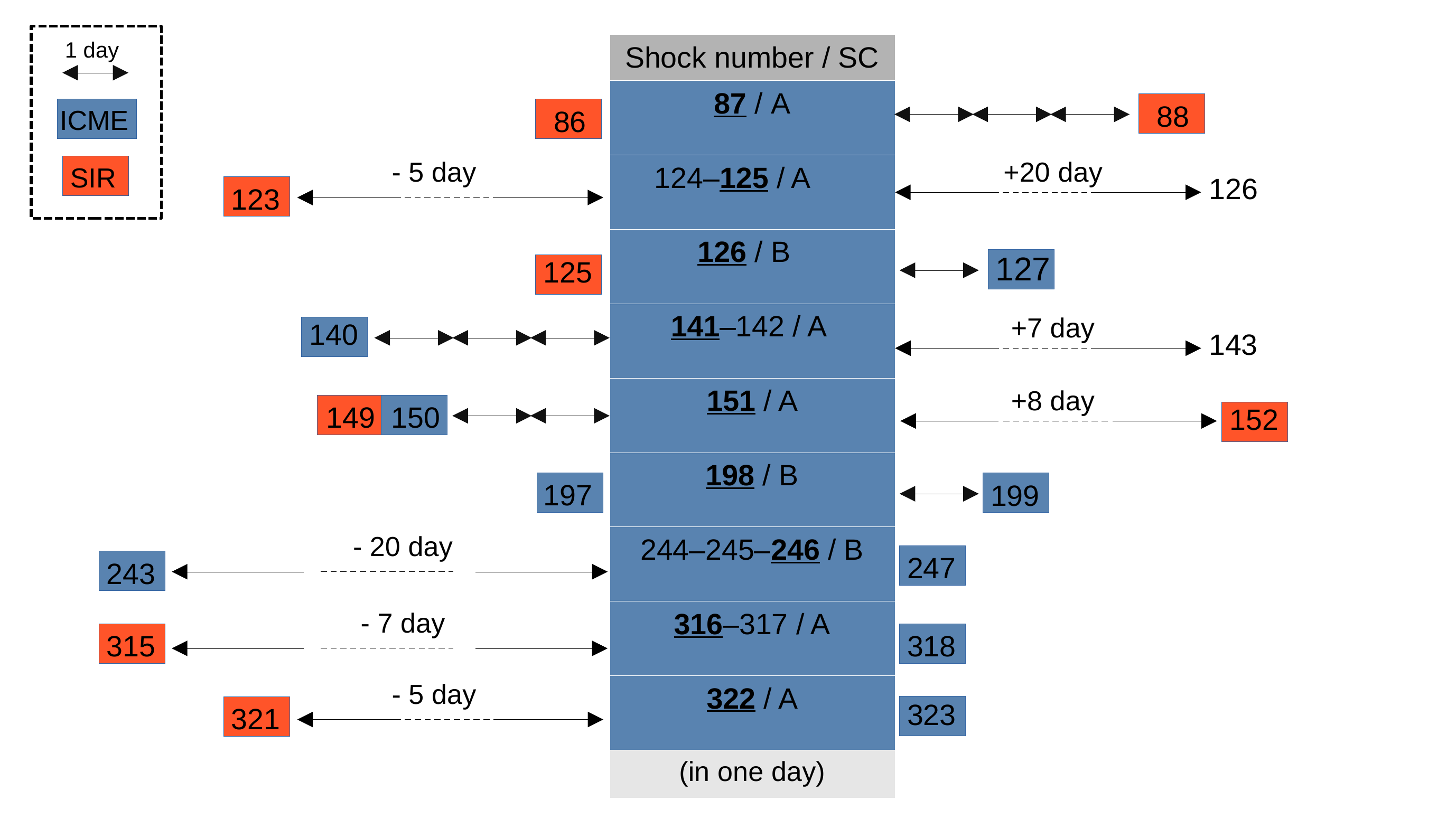}
      \caption{All the numbers represent the shock number from the STEREO list (\url{https://stereo-ssc.nascom.nasa.gov/pub/ins_data/impact/level3/IPs.pdf}). All of the ICME events in the blue central column occurred simultaneously with the relevant shock passage from our list of nine events. The nine numbers in bold print with underline indicate the events on our list. The SIR shocks from the list represented by the number of shocks inside the red box. Two way arrows indicate a day and $\mp$ day denotes a duration of more than a day between the occurrence of two shocks.  }
\label{fig:SIR}
\end{figure*}

All the events in Table \ref{table:1} are associated with interplanetary CME (ICME)-driven shocks. Having applied different window length from 15 to 480 minutes we found that the strongest signals in the NF data are usually found for large window length of 240 or 480 minutes.
In the two days before and two days after the shock passage time, we found other shocks associated with some of our events, which is illustrated in Fig. \ref{fig:SIR}. For example, shocks number 125 and 316 have one shock before the on the same day. The shock number 141 has one shock after, and shock 246 has two shocks before on the same day. In three events (shocks number 87, 126, 198), one shock occurred a day before; two of those are shocks related to stream interaction regions (SIRs). There was no previous shock in the five events (Fig. \ref{fig:SIR}). 

At the time of the in-situ shock crossing, radio spectrograms observed at the same STEREO spacecraft reveal another feature, which is present in all of the nine events: an increase in the dynamic spectrum simultaneous in a broad range of frequencies. The feature is most likely quasi-thermal noise extending from the kHz range up to $\sim$1 MHz. To gain a better understanding of this attribute, we investigated the STEREO list of shocks, and we found that this is a very common trait on the downstream of the shock crossing. Therefore, it is not a unique feature of the electron ESP events studied here.

In the following we will discuss some of the most interesting events in more detail.

\subsection*{Event on 28 May 2012 -- shock number 141}

This event (Fig.~\ref{fig:141}) is an example of a peak-like electron ESP event and it stands out from the others in terms of its very strong response in the e3 energy channel. LASCO-C2 recorded a fast Halo-CME with a linear speed of 1966 km/s at 2:24~UT on 27 May 2012.\footnote{\url{http://cdaw.gsfc.nasa.gov/CME_list}}

The STEREO/WAVES dynamic spectrum shows a type III radio burst on 26 May 2012 at 20:48 UT (third panel from top of Fig.~\ref{fig:141} (a)). In addition, a type II radio burst, a signature of a CME-driven shock accelerating electron beams \citep[e.g.,][]{ne85}, begins at 20:50 UT and continues until 23:20 UT between 16000-300 kHz. The first increase in particle intensities occurred simultaneously with the eruption on 26 May. A significant change of the intensity time profile occurs around 6:00 UT on 27 May 2012, when the fluxes begin to rise again after a continuous decrease and reach their maximum value 4 minutes before the shock time, in the upstream region. This is most likely due to the ICME-driven shock, whose contribution gets significantly stronger around this time (see Fig.~\ref{fig:141} near the black vertical dashed line marked with shock number 141). 

The obliquity $\theta_{Bn}$ of the in-situ shock is almost 75° and the Mach number is 2.85. The NF data of the electron channels utilizing various window lengths are displayed in the six lower panels of Fig.~\ref{fig:141}. Significant changes in the SEP intensities start to show up three hours before the shock, and the fluxes of all energy channels starts to rise, which also appears as signal changes in the NF data. The signal in the e3 channel is the strongest compared to the other electron energy channels (blue line in Fig.~\ref{fig:141} plots) and even greater than those of the proton channels (Fig.~\ref{fig:141} (c)), despite the fact that protons have a higher intensity than electrons in the time-intensity profile (top panel). 

Between the other windows lengths, the highest value of the z-score at the time of shock crossing belongs to $M = 480$. 
The energetic particle time profiles in the top panel of Fig.~\ref{fig:141} show an intensity decay immediately after the shock crossing, observed in both electrons and protons.
The magnetic field variations in this event indicate that the ICME driving the shock and related to the CME associated with the SEP event has a magnetic cloud-like structure with a gradually rotating and enhanced-in-magnitude magnetic field. However, the field shows a strong depletion in magnetic field magnitude around 9:00 UT, which may indicate the presence of a pre-existing ICME ahead of the driving ICME that is related to the fast halo CME. The sheath region of the driving ICME has a complex magnetic structure and could even have a partly closed magnetic topology. This, however, needs a more thorough dedicated study in the future.


\subsection*{Event on 23 July 2012 -- shock number 151}

This event (Fig.~\ref{fig:151}) has been studied in detail by several authors \citep[e.g.,][]{Russel2013,Temmer2015,Riley2016} as it is one of the most energetic eruptions observed during the space era. We do not provide a detailed analysis of this event but examine the key observational properties in relation to our event sample. The event is an example of a plateau + peak-like shaped electron ESP event, with electron fluxes higher than proton fluxes for the majority of the time-intensity profile. 
The LASCO CME catalog reports a fast CME with a speed of 2003 km/s at 2:24 UT on 23 July 2012, followed by a clear type II radio burst (third panel from top Fig.~\ref{fig:151} (a)), which continued until 21:40 UT visible by both STEREO A and B (not shown). HET measured a very gradual SEP event associated with a type III burst at 01:45 UT on 23 July 2012 . The enhancement of particle intensities associated with the fast forward ICME-driven shock observed by STEREO~A at 20:55 UT on 23 July 2012 reached its peak value 1 minute before the shock time, in the upstream region. It is an oblique shock with $\theta_{Bn}=45^\circ$ and a Mach number of 2.46 based on the shock list (but we note that many of the earlier detailed analyses give somewhat different values for the obliquity and strength of the shock). The electron flux increases four orders of magnitude during this event making it an exceptionally intense event. By zooming in around the ESP event (Fig.~\ref{fig:151} (b) and (c)), we see a pronounced plateau-shaped intensity increase between 19:15 to 22:56 with the time of the shock crossing in the middle. During this time, the magnetic field gradually starts to weaken and only a small jump in magnetic field magnitude is observed at the shock (2nd panel from top).
In the NF data, the HET e3 channel (2.8--4.0 MeV) exhibits the strongest shock response.
The $M = 480$ window length had the highest value by computing the z-score around the shock crossing in six different window lengths.
The end of the plateau-like shape of intensities around 23:00 UT on 23 July is accompanied by the beginning of an ICME, which has a highly compressed magnetic field reaching values beyond 100 nT. The ICME has again a complicated structure that may consist of two individual flux ropes \citep{Russel2013, Riley2016}, first of which would be related to a pre-existing CME in the solar wind overtaken by the fast eruption. This implies that the strong decrease of fluxes at the onset of the ICME would be related to a change in the topology of the field.

\subsection*{Event on 6 November 2013 -- shock number 198}

Shock number 198 in Fig.~\ref{fig:198}, which occurred on 6 November 2013, is unique in terms of being the only MeV electron ESP event in our sample without a previous SEP event \citep{Chiappetta_2021}. According to the SOHO LASCO CME Catalog, the associated fast Halo-CME occurred with a linear speed of 1040 km/s at 5:12~UT on 4 November 2013. Both STEREO/WAVES recorded a strong type II radio burst extending to very low frequency (not shown). Magnetic field components increased in the first hours of 5 Nov 2013 in the upstream region at the same time as the quasi-thermal noise signature in the radio spectrogram. About 40 minutes before the shock crossing a small type III radio burst appeared in the radio spectrogram. This new solar event might have injected SEPs at the right time to provide seed particles for the in-situ shock on 6 Nov leading to electron intensity peaks close to the shock crossing time (31 min) on the downstream side.
The e2 and e3 channels responded in a similar way as indicated by the NF data, but the response of the e1 channel is weaker. The magnetic field associated with the ICME driving this event is less intense than in the other two example events. It is characterized by the presence of large-amplitude fluctuations throughout the event. Also in this case, the downstream of the shock seems to host a compressed magnetic field region between the shock and the driving ICME related to the fast halo CME (from about 02:40 to 05:00 UT), which coincides with the start of the decrease of particle fluxes indicating a possible change in field topology. The structure of the field and the duration of the compressed region does not, however, seem to be consistent with a double flux-rope structure like in the 23 July 2012 event.

\subsection*{Shock number 199 -- example of an event missed by the filtering method}
Shock number 198 is followed by a gradual SEP event that occurred on 7 Nov 2013 (see Fig.~\ref{fig:199} (a)) in conjunction with a Halo CME, resulting in a type II radio burst in the dynamic spectrum. The SEP event was also studied by \citet{Dresing2016b}, who reported that STEREO~B was situated inside a magnetic cloud of a pre-event ICME during the start of the SEP event. 

The 7 Nov 2013 CME drove a shock that passed STEREO~B on 8 Nov (shock number 199). The SEP event has a long duration and a plateau-type profile, also in electron channels e1--e3. This could be an indication of a prolonged electron injection that could be related to continuous acceleration at the shock. However, the signal of the NF data does not show a significant response of the electron channels for the associated IP shock crossing of shock number 199 (Fig.~\ref{fig:199} b). This implies that the shock does not maintain its ability to accelerate electrons up to 1 AU. 

\begin{figure*}
    \centering
    \subfigure(a){\includegraphics[width=0.312\textwidth]{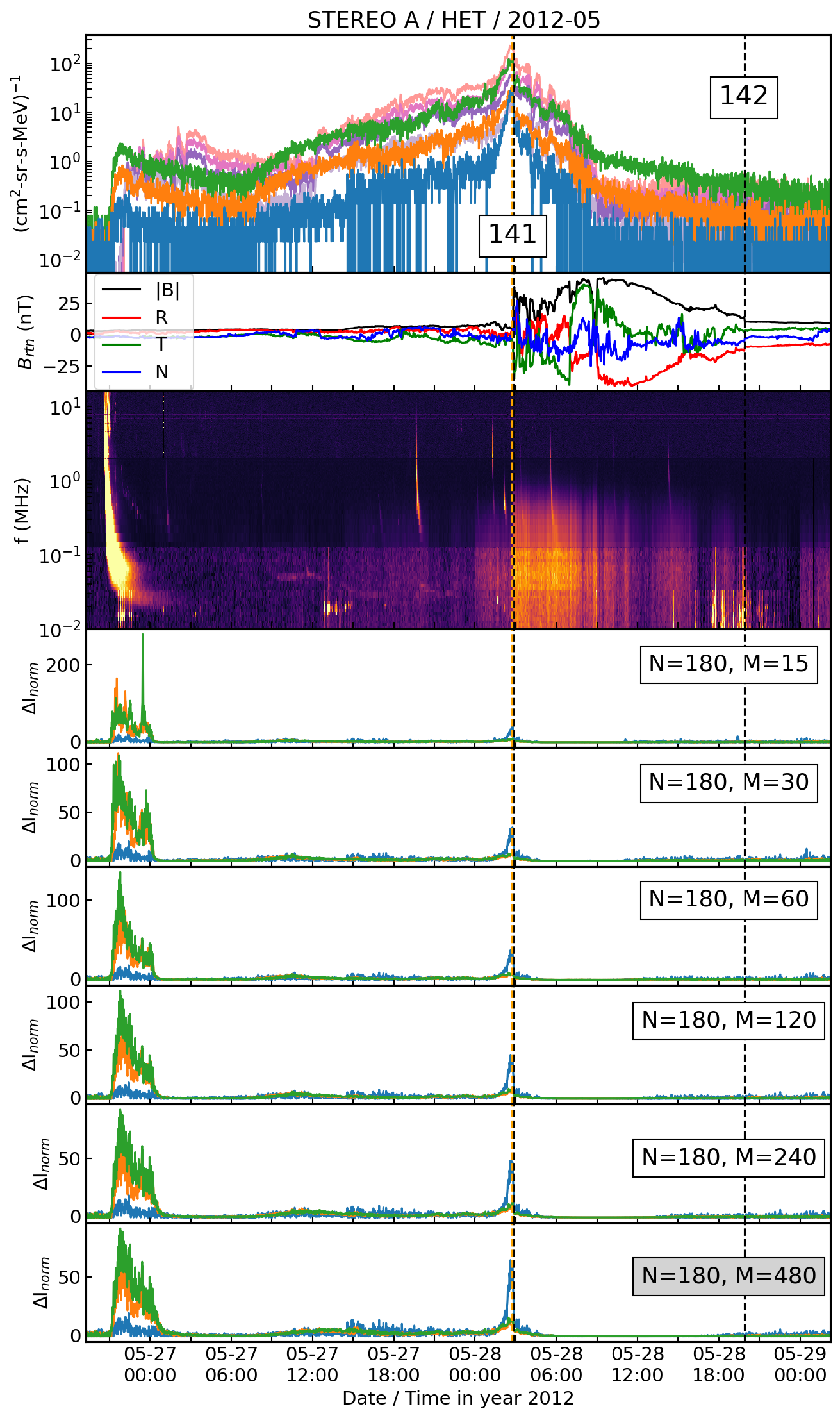}}
    \subfigure(b){\includegraphics[width=0.30\textwidth]{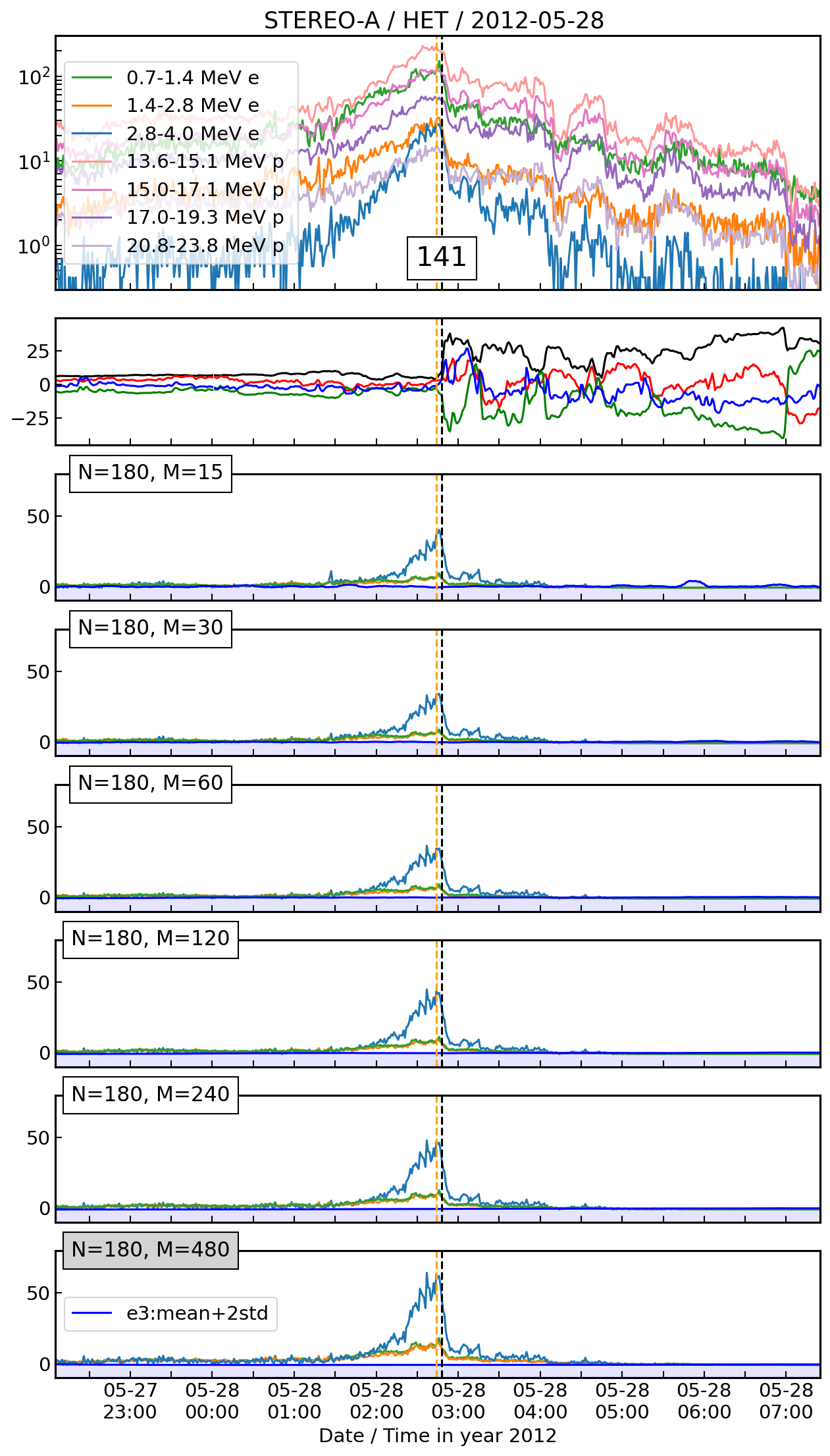}}
    \subfigure(c){\includegraphics[width=0.312\textwidth]{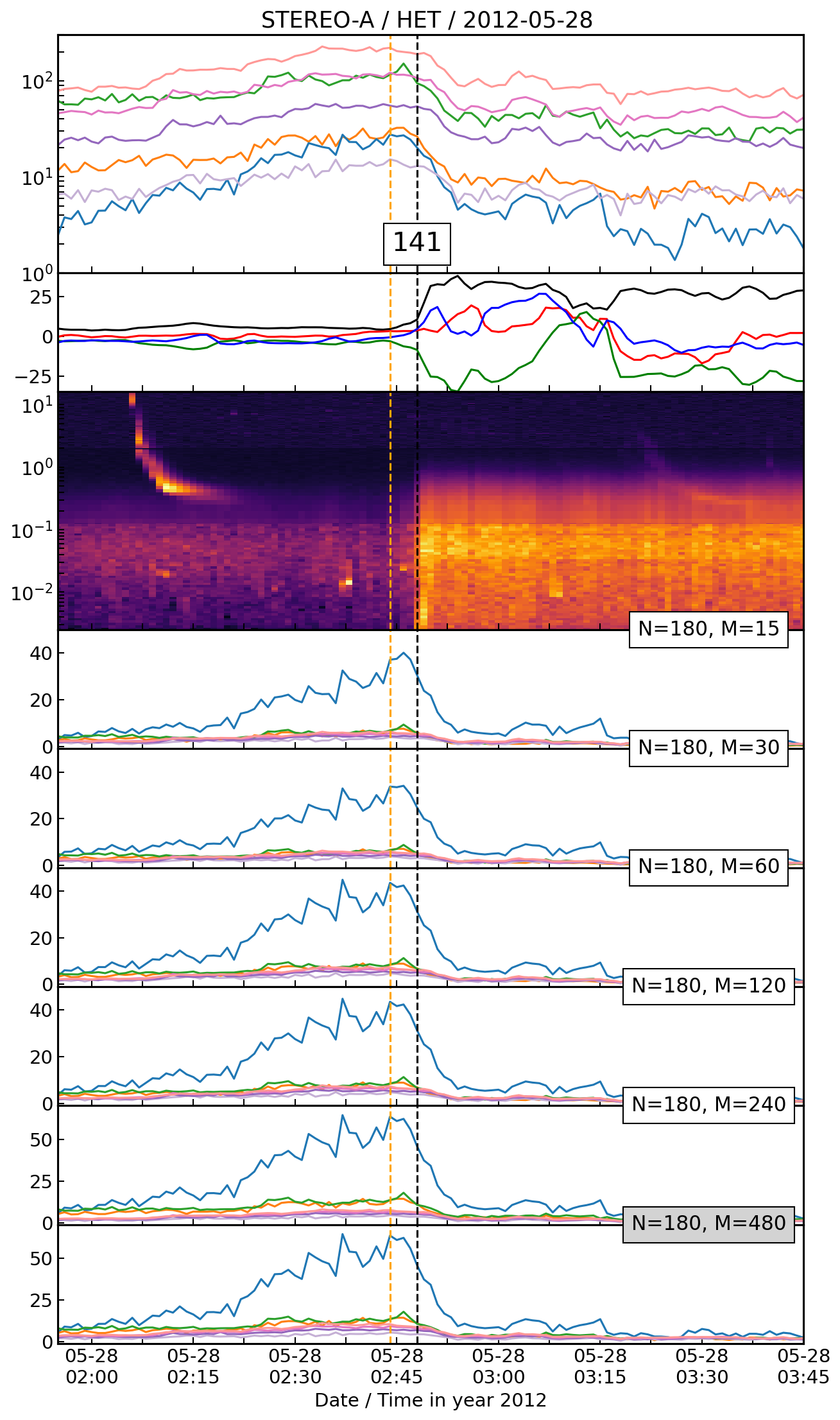}}
    \caption{SEP event on 27 May 2012 and corresponding ESP event on 28 May 2012 observed by STEREO~A. (a) \textit{from top to bottom}: time profile of proton and electron intensities at different energy channels as indicated in the legend in the top panel of (b), the evolution of magnetic field magnitude and its components in RTN coordinates, STEREO/WAVES dynamic spectrum. The lower six panels show the normalized filtered data of the three electron channels using a fixed time lag parameter of $(N = 180)$ and increasing window lengths $M$. Figures (b) and (c) are the same format as \ref{fig:125}. The boxes show the N and M value of each panel and the gray one indicates the the one yielding the most significant signal of the NF data.}
    \label{fig:141}
\end{figure*}
   

\begin{figure*}
   \centering
   \subfigure(a){\includegraphics[width=0.318\textwidth]{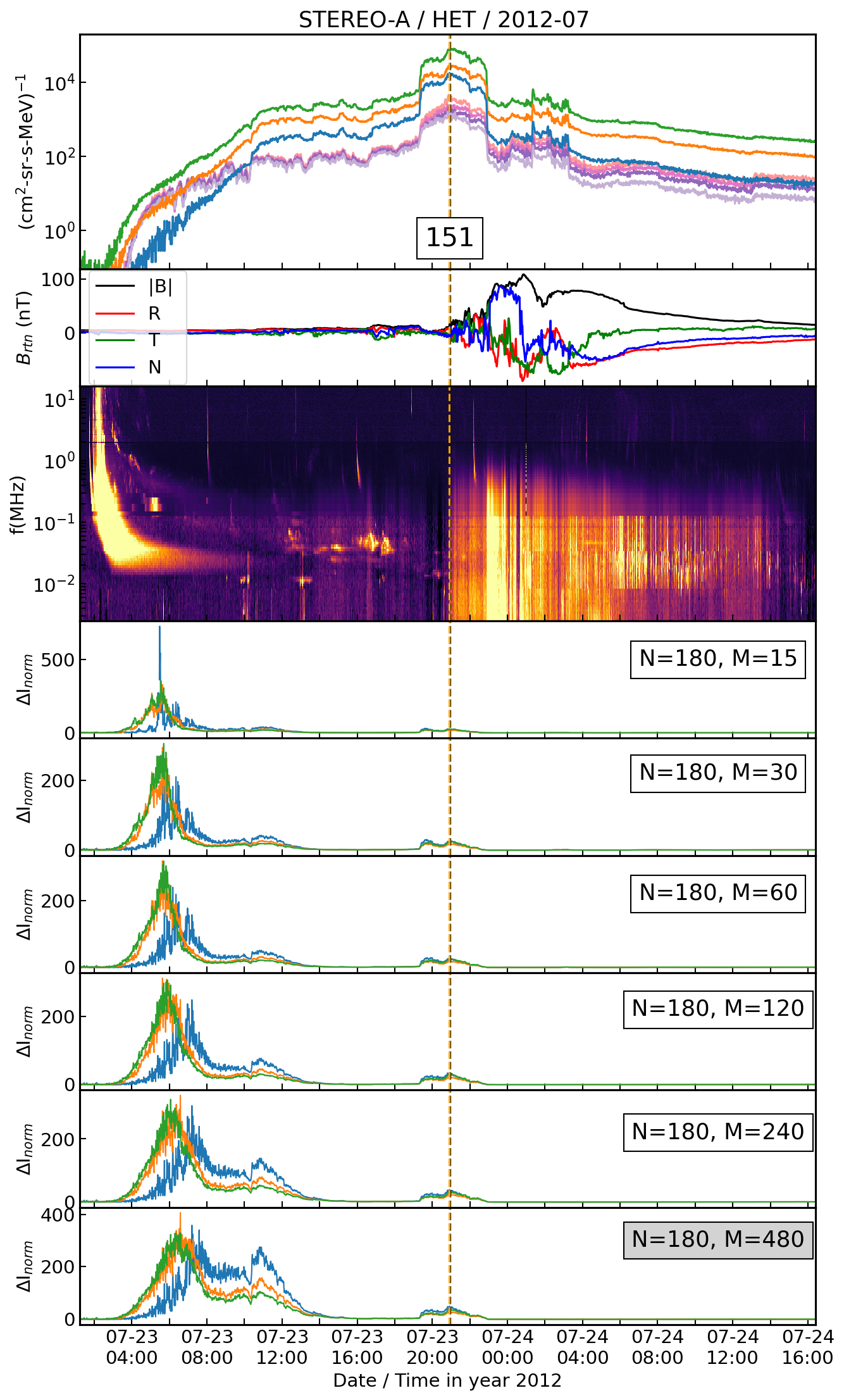}}
   \subfigure(b){\includegraphics[width=0.30\textwidth]{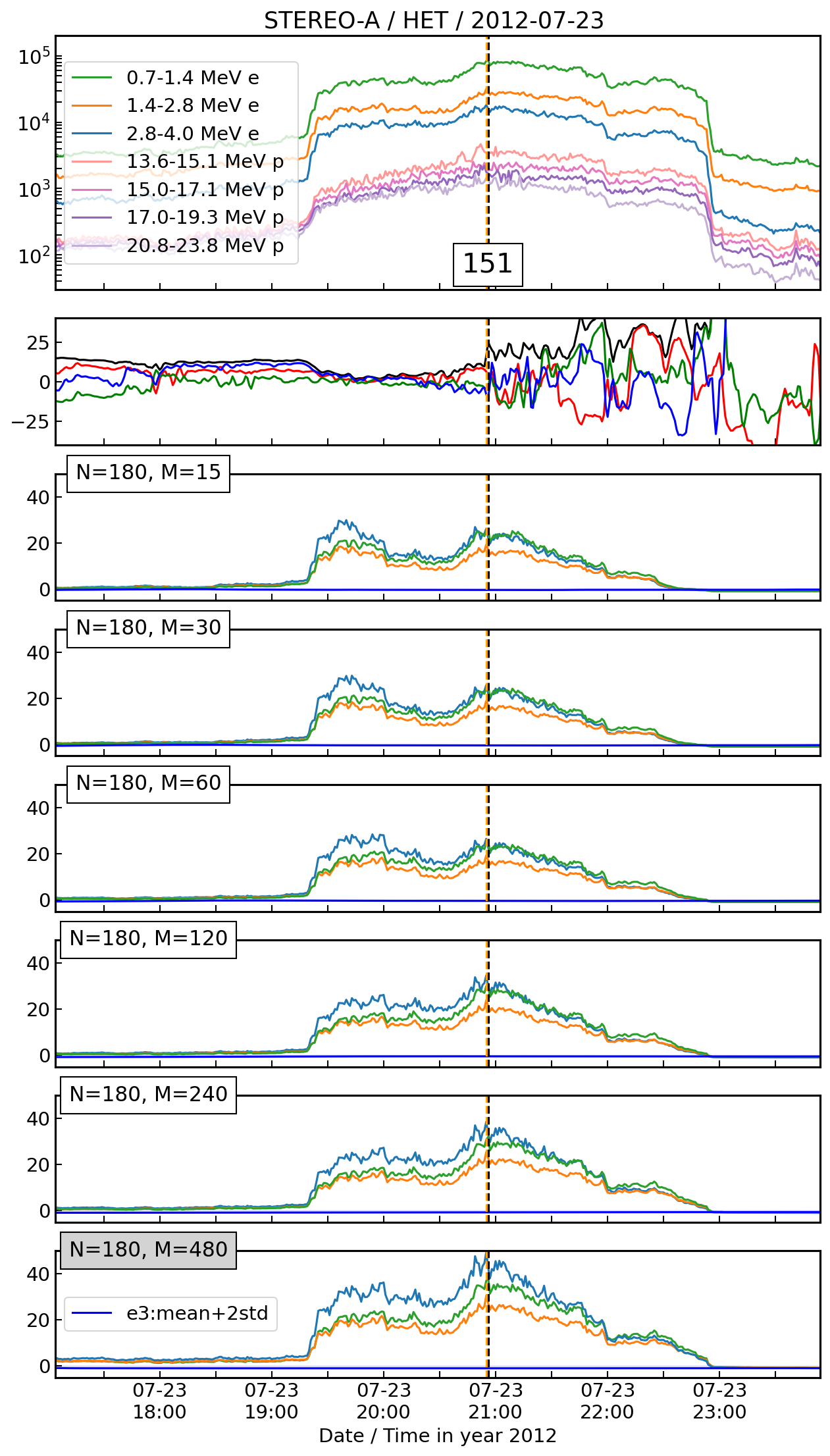}}
   \subfigure(c){\includegraphics[width=0.308\textwidth]{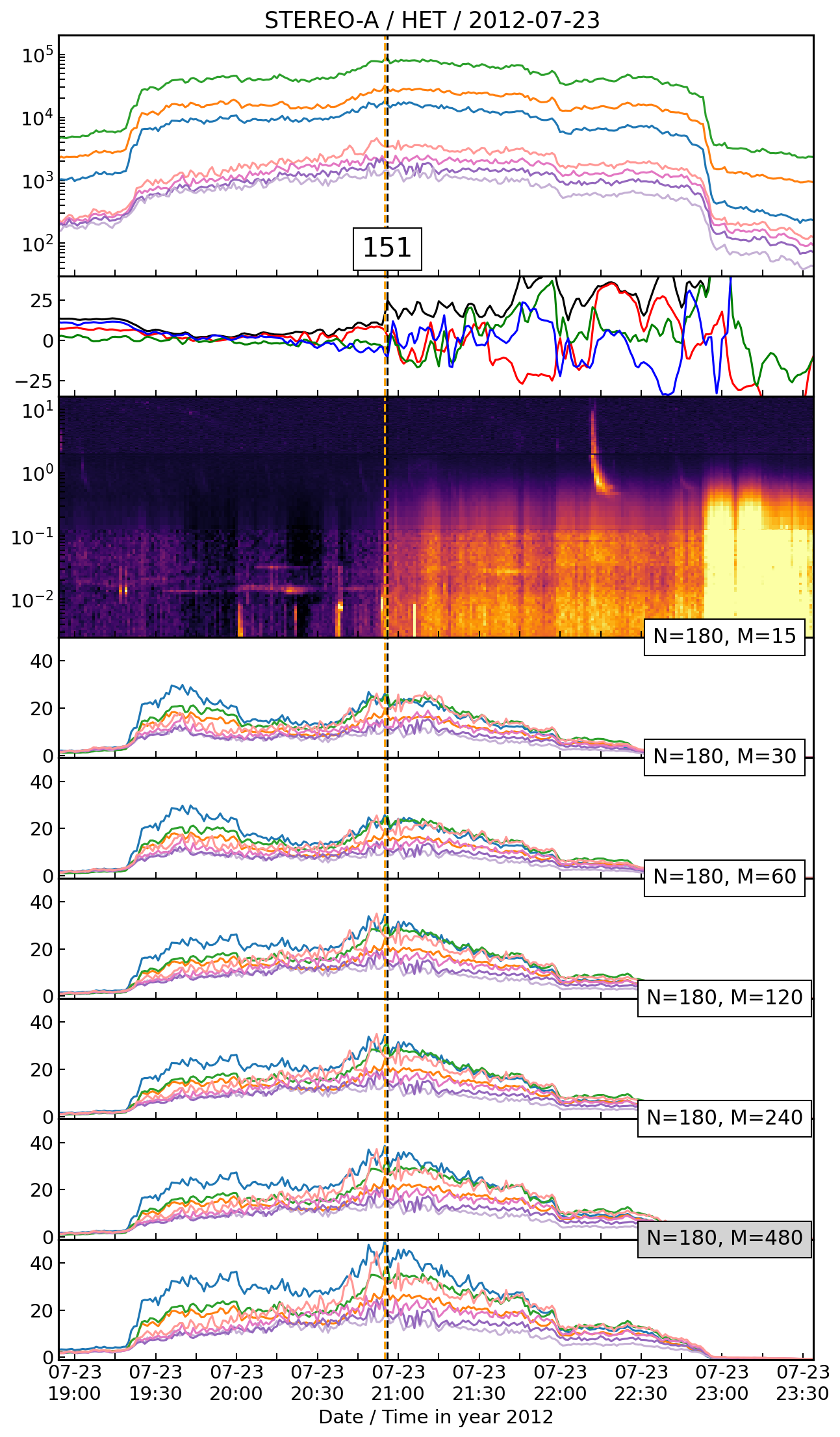}}
    \caption{Summary of electron and proton observations by the STEREO/HET instrument on 23 July 2012. Figures (a), (b) and (c) have the same format as Fig.~\ref{fig:141}}
    \label{fig:151}
\end{figure*}

\begin{figure*}
\centering
\includegraphics[width=0.31\textwidth]{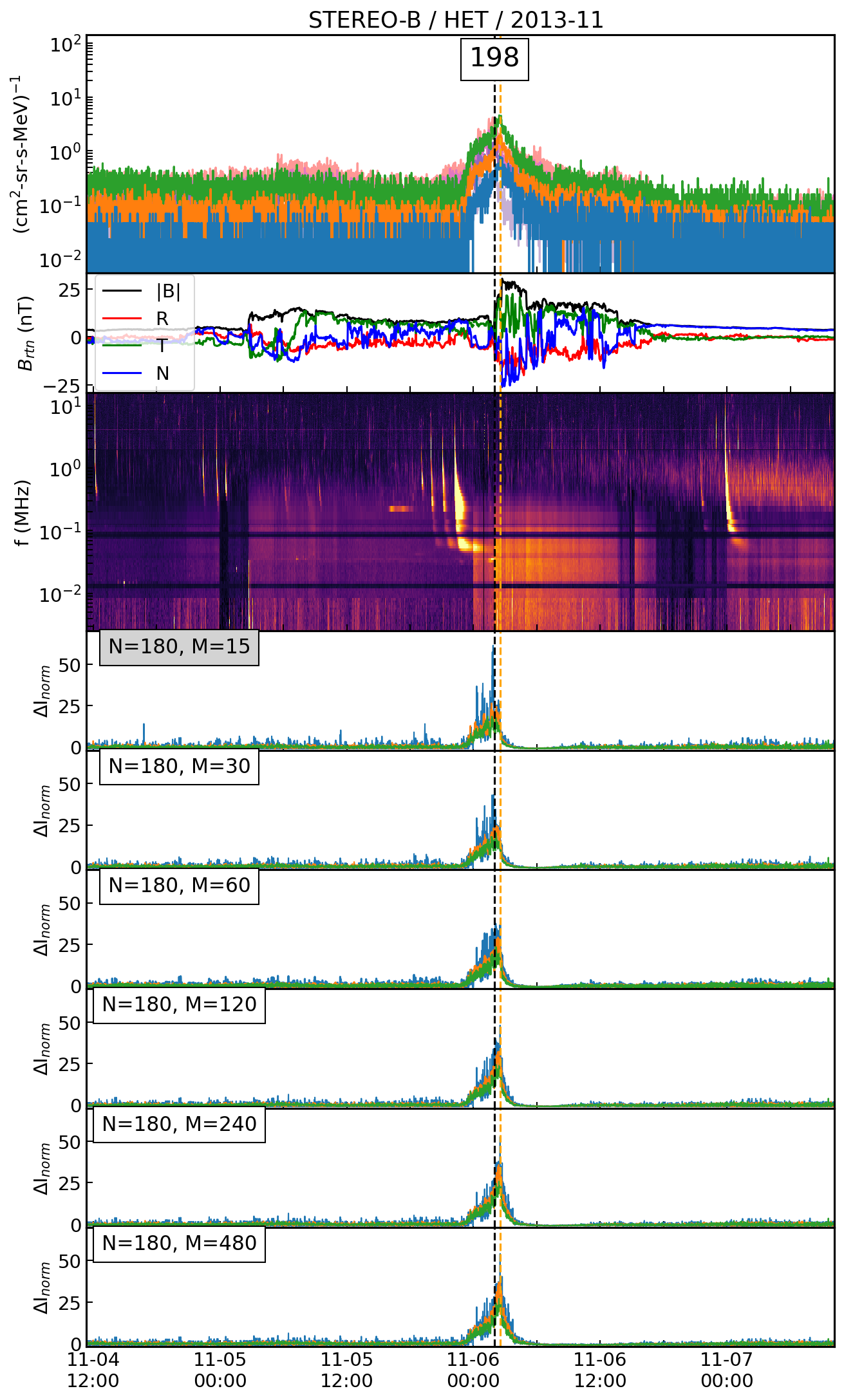}
\includegraphics[width=0.30\textwidth]{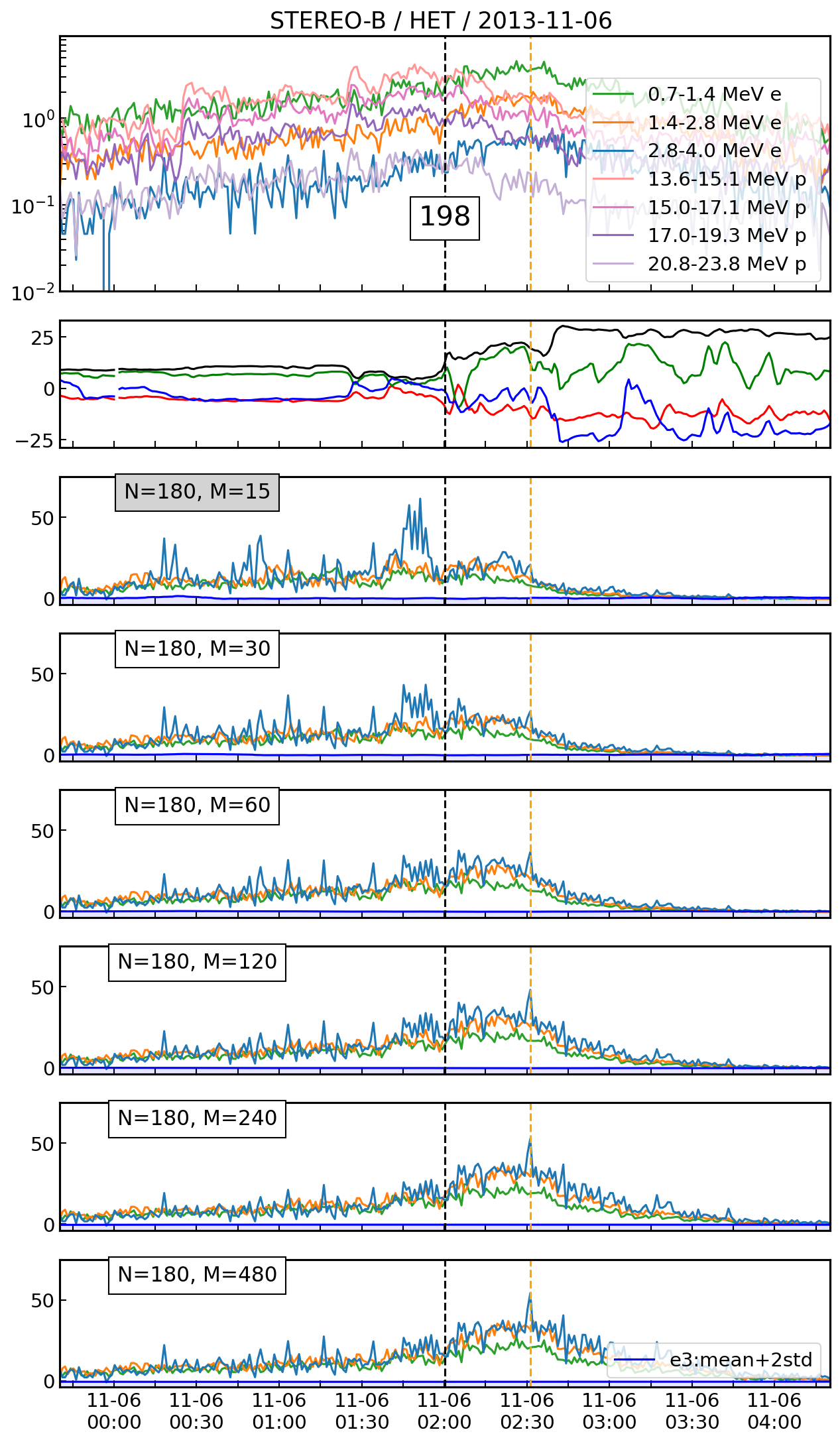} 
\includegraphics[width=0.30\textwidth]{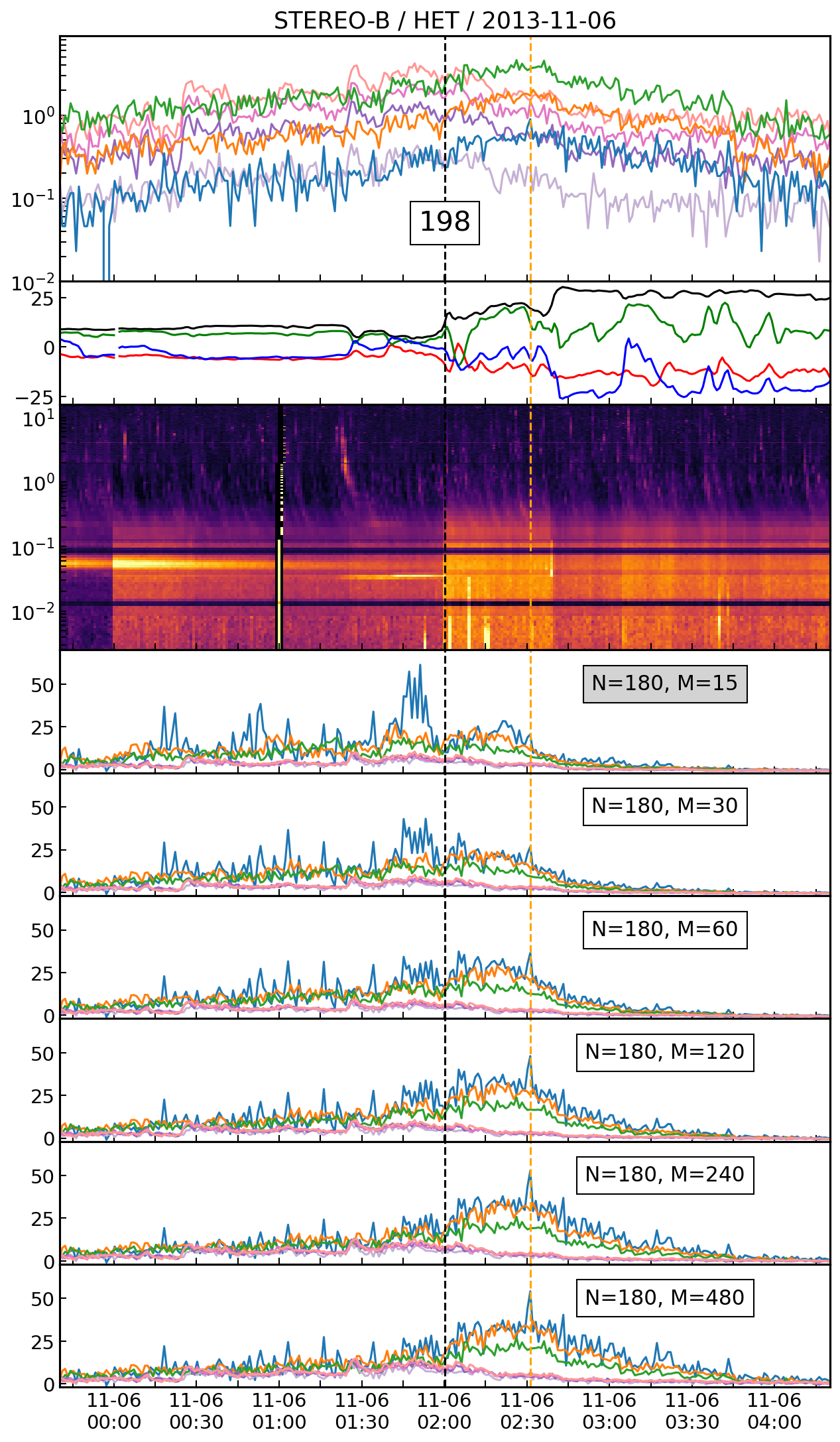}
\caption{Same format as Fig.~\ref{fig:141} showing STEREO~B/HET observations of the event associated with the ICME-driven shock on 6 Nov 2013.}
\label{fig:198}
\end{figure*}

\begin{figure*}
\centering
\includegraphics[width=0.309\textwidth]{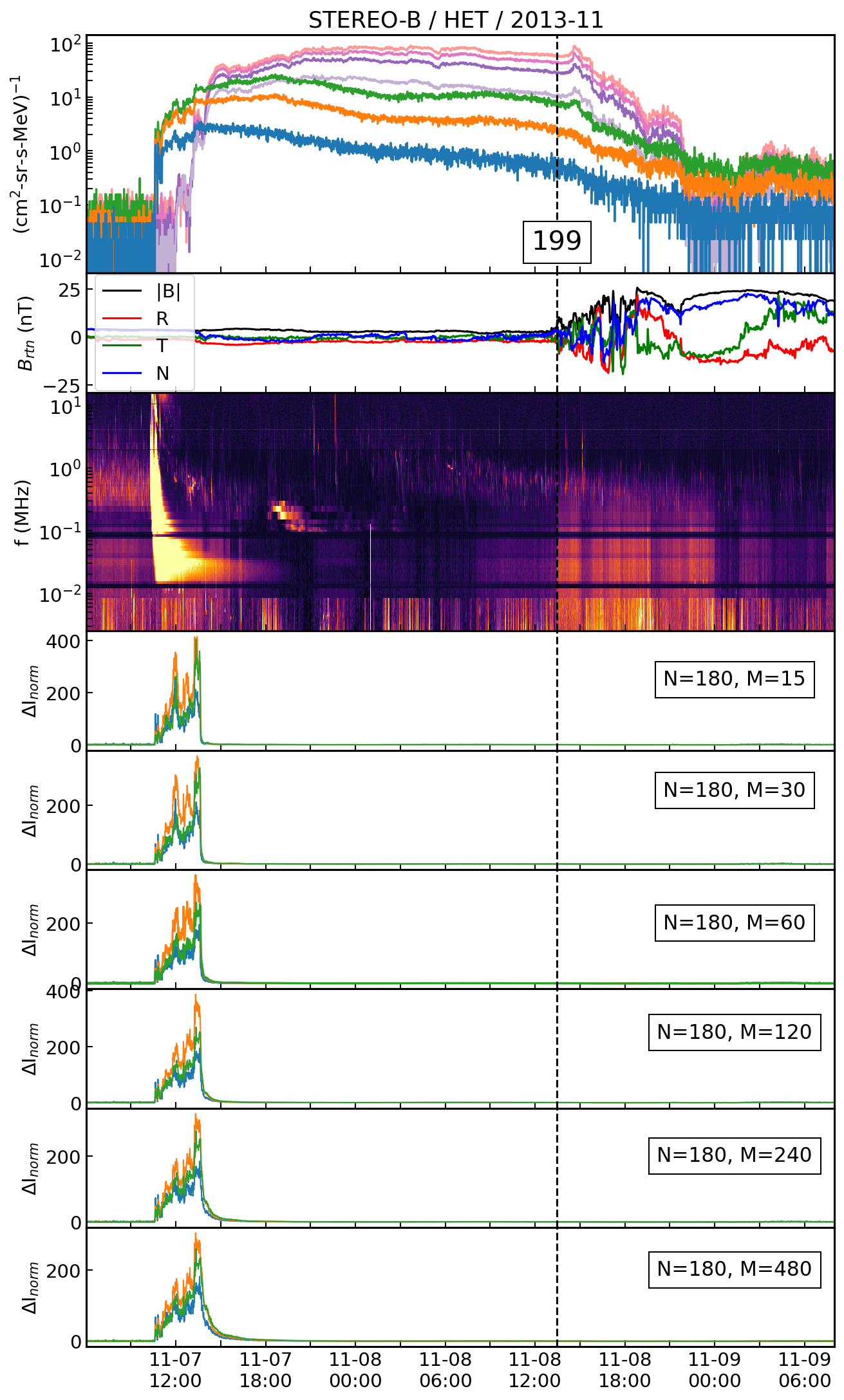}
\includegraphics[width=0.30\textwidth]{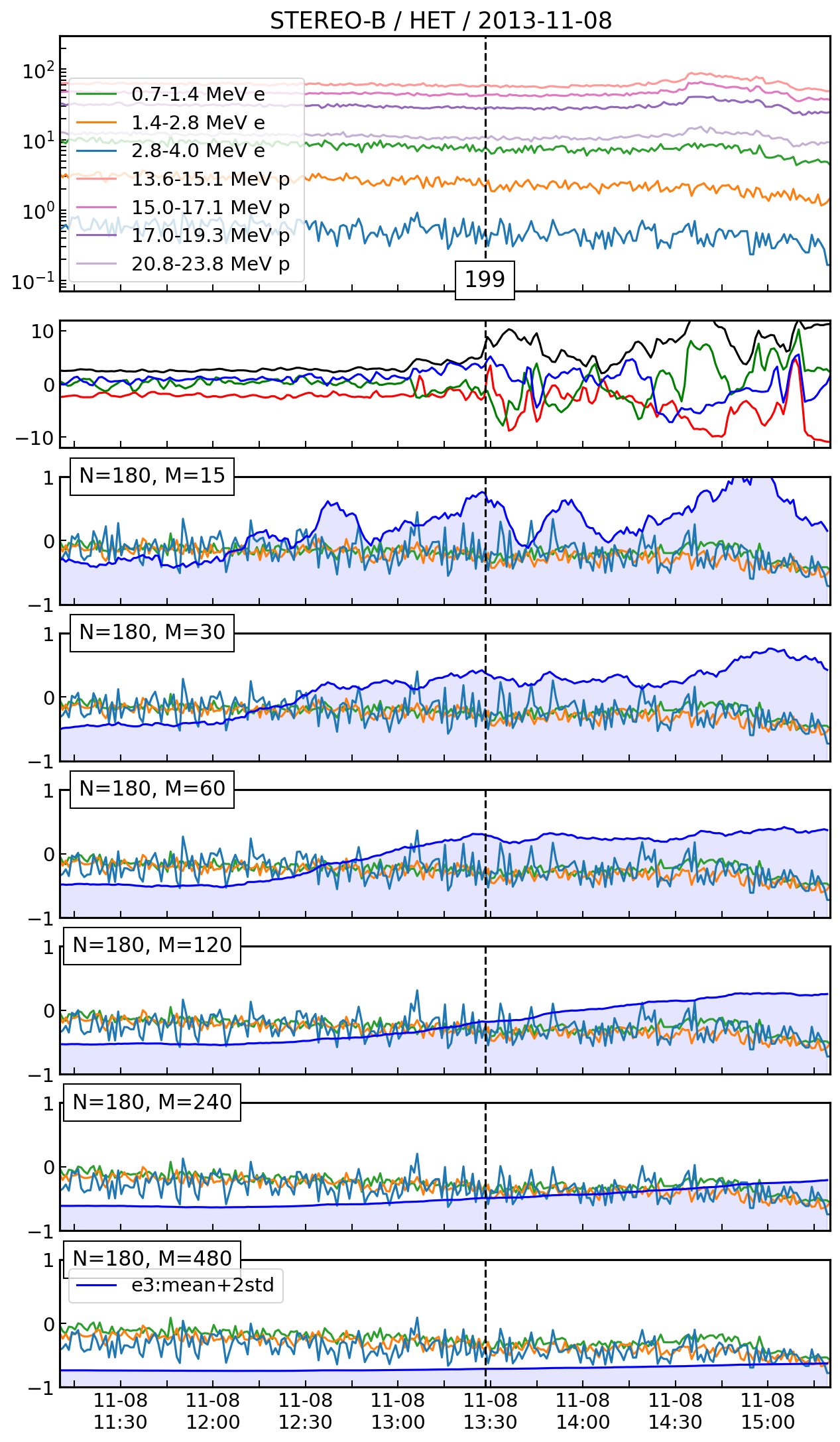} 
\includegraphics[width=0.30\textwidth]{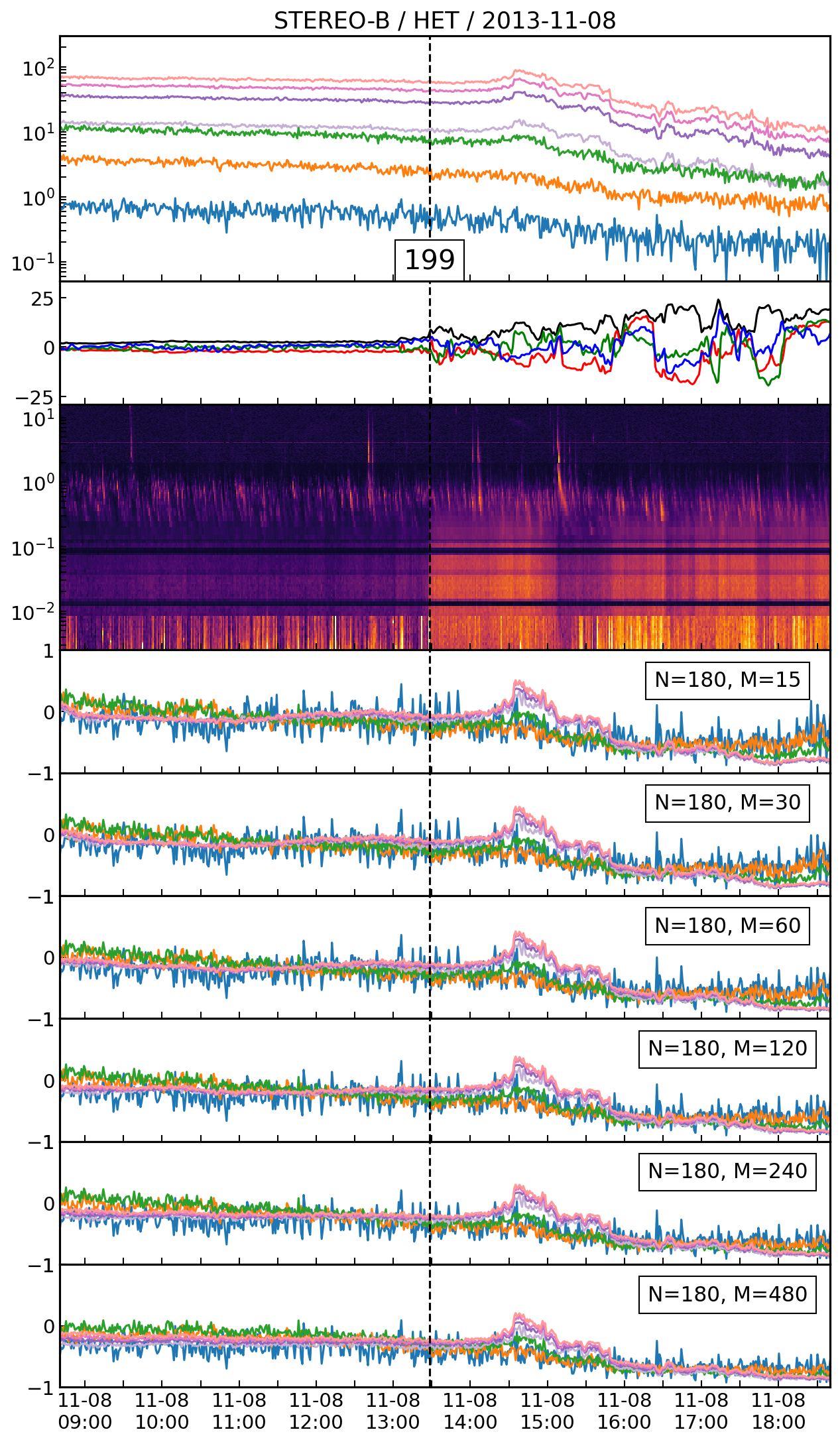}
\caption{Same format as Fig.~\ref{fig:141} but showing STEREO~B/HET observations the event associated with the ICME-driven shock on 8 Nov 2013.}
\label{fig:199}
\end{figure*}
\section{Discussion and Conclusion}\label{sec:discon}
We scanned the STEREO mission for interplanetary shocks that show local acceleration signatures of relativistic electrons. Nine events were identified by applying a novel filtering method that identifies significant peaks in time-intensity profiles of HET energy channels. All shocks were associated with ICMEs. The events were divided in three classes based on their time-intensity profiles, i.e., plateau-like, peak-like and plateau+peak-like events. The identified electron ESP events had variable electron to proton ratios, some being proton dominated while some were dominated by electrons in the STEREO/HET energy range.

The nine events that show electron acceleration signatures are all related to very fast shocks with transit speeds exceeding 900~km/s. The transit speed was recently identified by \citet{Ameri2022} as the best correlated shock property with the $\sim$10 MeV proton intensities at the shock in ESP events. A high transit speed may imply that a shock has been strong all its way from the Sun to the observer, allowing it to maintain its acceleration efficiency for an extended amount of time. 

Three ESP events were analyzed in more detail. Two of them were associated with SEP events and one was an isolated ESP event. The ones associated with SEP events were more intense. Their associated ICMEs had complex magnetic structure with indications of interactions in the interplanetary space between pre-existing ICME structures and the fast eruption driving the event. The isolated ESP event is not as strong as the ones associated with SEPs. It, too, had a complex ICME driver, although the peak magnitude of the interplanetary magnetic field was not as high as in the SEP-event-related events. Although it lacked an SEP event related to the eruption, the isolated event had a type III burst occurring a few hours before the shock passage just before the start of rise of the energetic particle fluxes. This solar event could have provided seed particles for the shock and thus facilitated the acceleration of electrons to energies detected in HET energy range.

While it performs well with ESP-like events, our filtering method is unable to detect more gradual contributions of shocks to the energetic electron fluxes. Some events (e.g., shock no. 199) show plateau-like but gradually decaying intensities until the shock passage, and then faster decays after the passage of the leading edge of the ICME. In such cases, shocks may still have significantly contributed to electron acceleration as the shock propagates from the corona to 1 AU. Equipped with the new instruments onboard Solar Orbiter and Parker Solar Probe that reach close distances to the Sun, such events may appear as much more clear-cut ESP events. 

In conclusion, while ESP events accelerating electrons to relativistic energies are not very common at 1 AU, there are still a number of cases that show signatures of local acceleration by the shock and/or its sheath region up to several MeVs. The next solar maximum with fast CMEs and new observational capabilities will likely bring more clarity to the issue of electron acceleration by interplanetary shocks. 

\begin{acknowledgements}
We acknowledge financial support from the European Union’s Horizon 2020 research and innovation programme under grant agreement No.\ 101004159 (SERPENTINE) and the support of Academy of Finland (FORESAIL, grants 312357 and 336809). 

SOHO LASCO CME catalog is generated and maintained at the CDAW Data Center by NASA and The Catholic University of America in cooperation with the Naval Research Laboratory. SOHO is a project of international cooperation between ESA and NASA.

\end{acknowledgements}

%
%

\bibliographystyle{aa}
\bibliography{ref}

\begin{appendix}
\section{Additional STEREO/HET observations of electron ESP events}\label{app:plots}
For each event listed in Table \ref{table:1} we show a figure in the format of Fig.~\ref{fig:141} presenting a long duration time profile of the SEP event and corresponding ESP particle intensities measured by STEREO/HET, the magnetic field magnitude and its components measured by the IMPACT/MAG instrument in RTN coordinates and a STEREO/WAVES dynamic spectrum. We also show the normalized filtered data with the fixed value of three-hour time lag ($N$) for all the cases and six given values for a temporal window of background intensity from 15 minutes to 8 hours ($M$). The figures contain the zoom-in profile around the shock crossing time and the shaded area of the mean plus two times of SD of the moving background window (the color of the shaded area is based on the electron energy channel showing the strongest signal in terms of of NF data). In addition, NF data for the proton channels have been added for comparison. The black and orange dashed lines on the plots represent the time of shock crossing and peak intensity time of selected energy channel of electron, respectively.
\begin{figure*}
\centering
\includegraphics[width=0.32\textwidth]{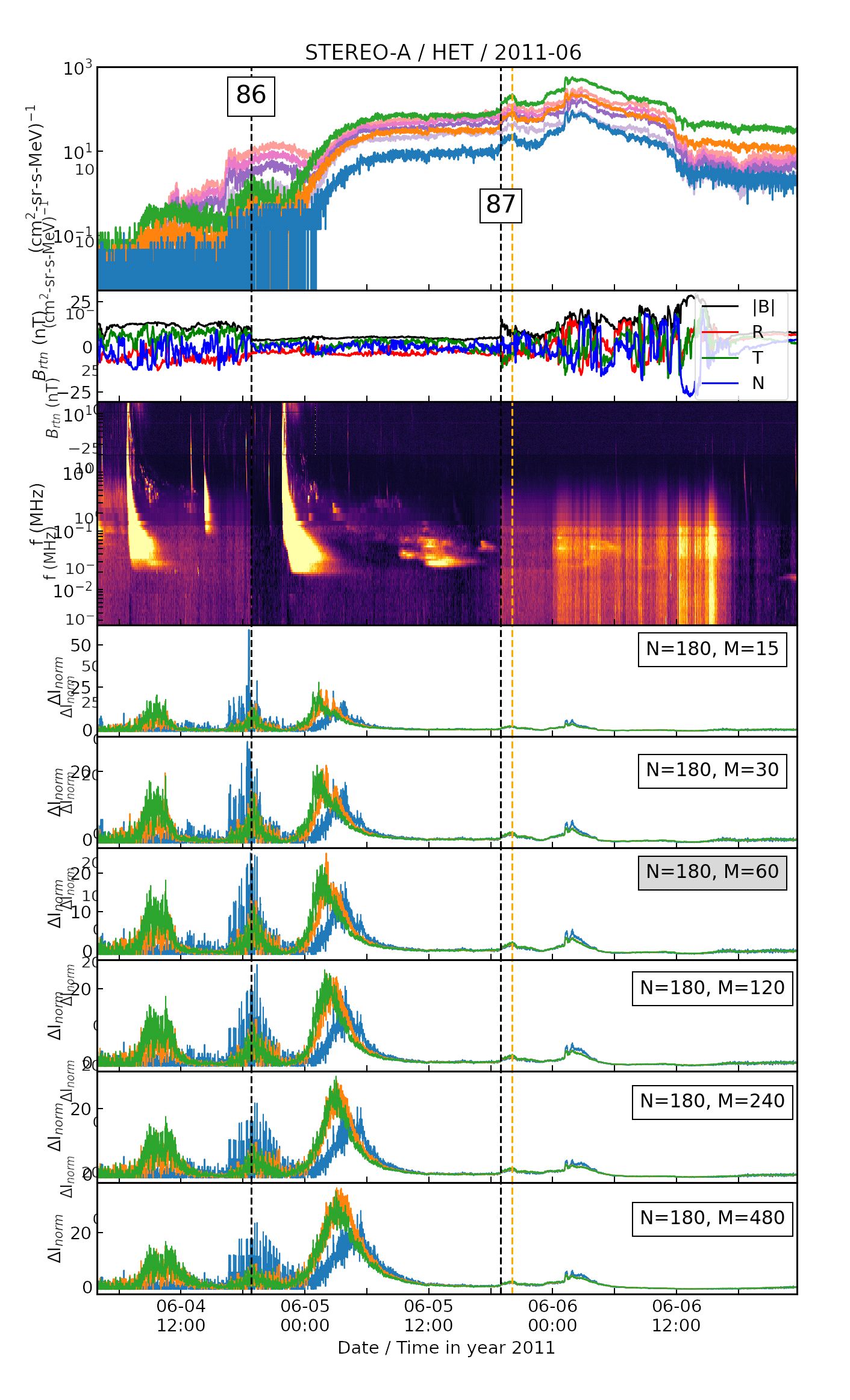}
\includegraphics[width=0.30\textwidth]{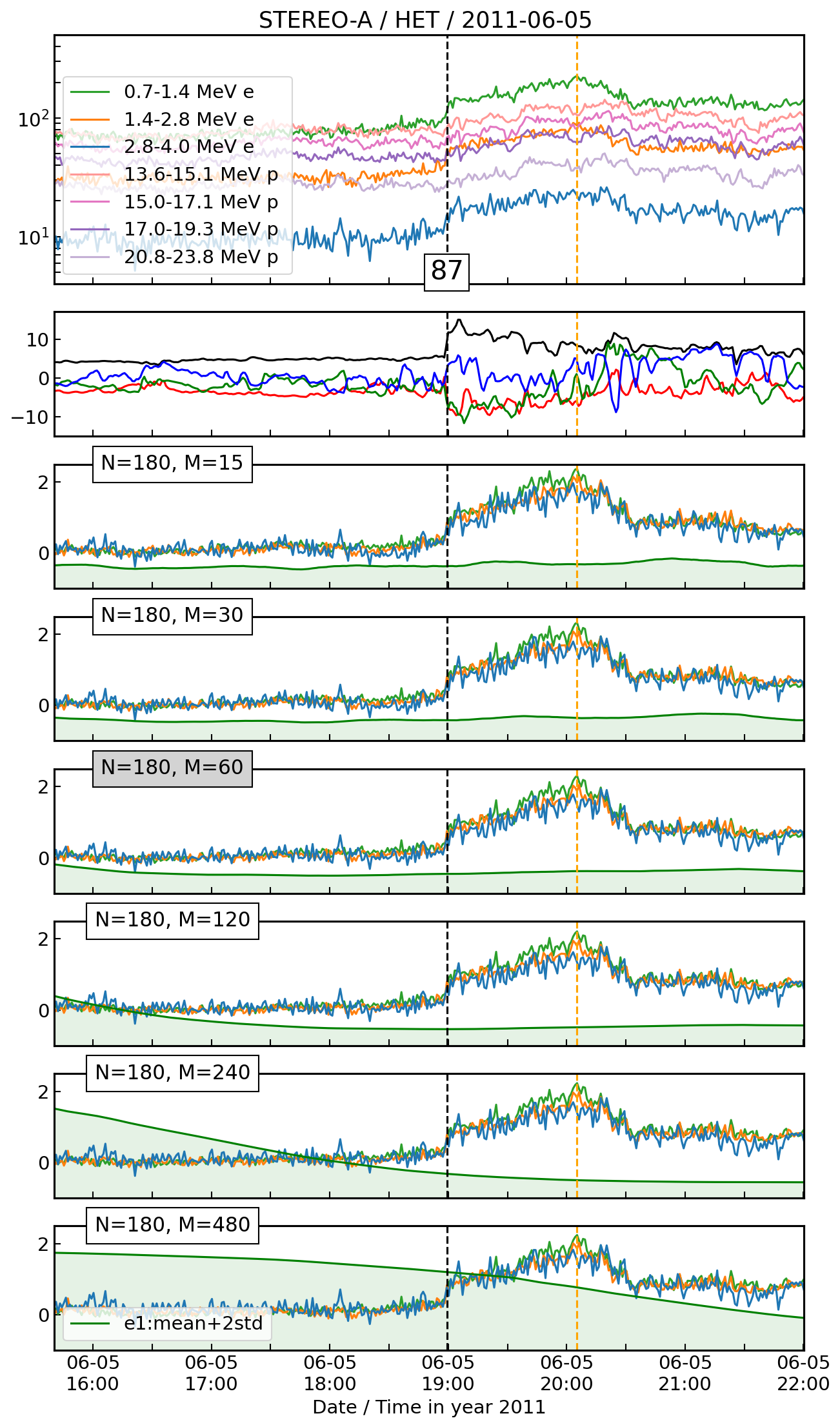} 
\includegraphics[width=0.30\textwidth]{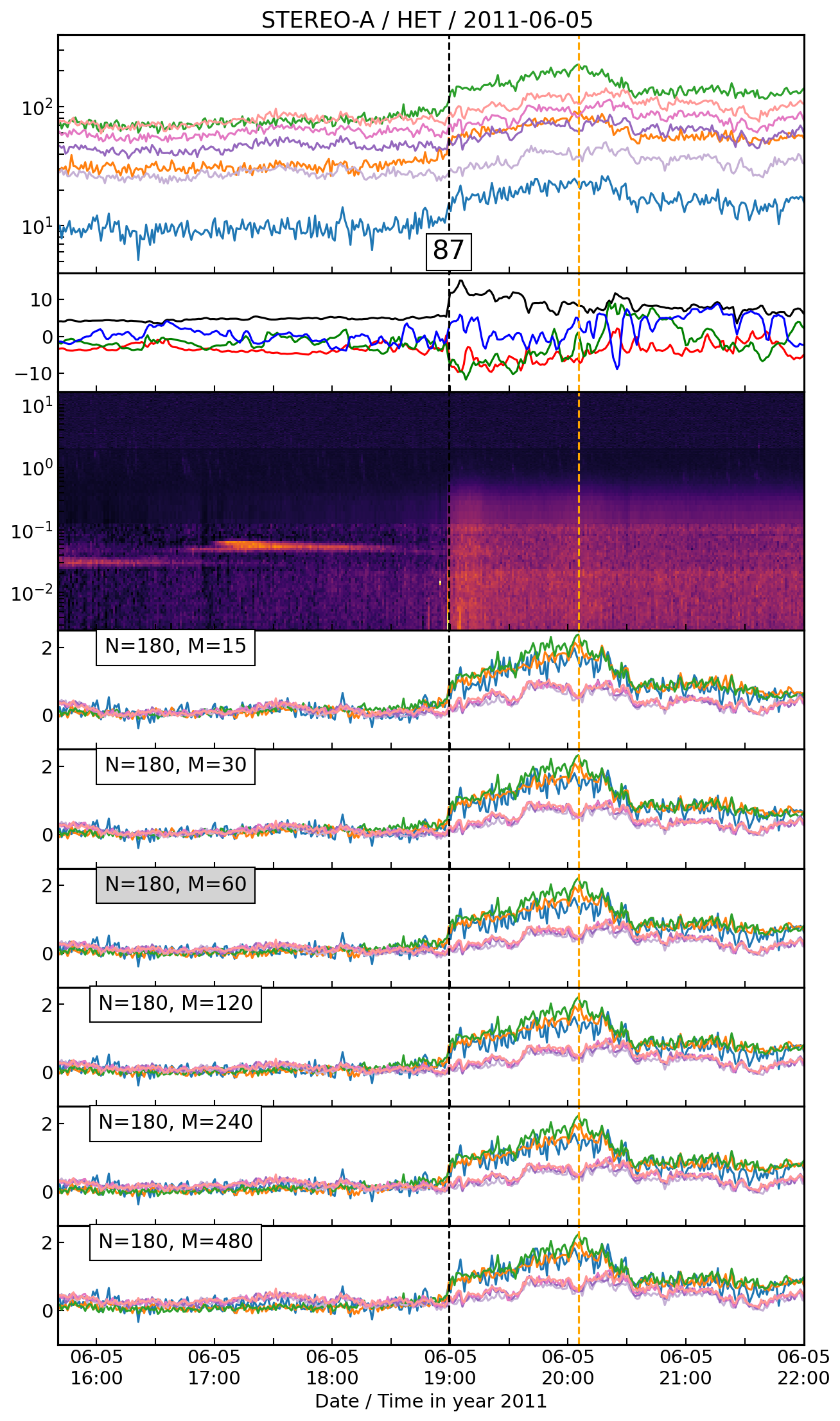}
\caption{Same format as Fig.~\ref{fig:141} but showing STEREO~A/HET observations  of an SIR-associated shock on 4 June 2011 and a ICME-driven shock on 5 June 2011.}
\label{fig:87}
\end{figure*}

\begin{figure*}
\centering
\includegraphics[width=0.32\textwidth]{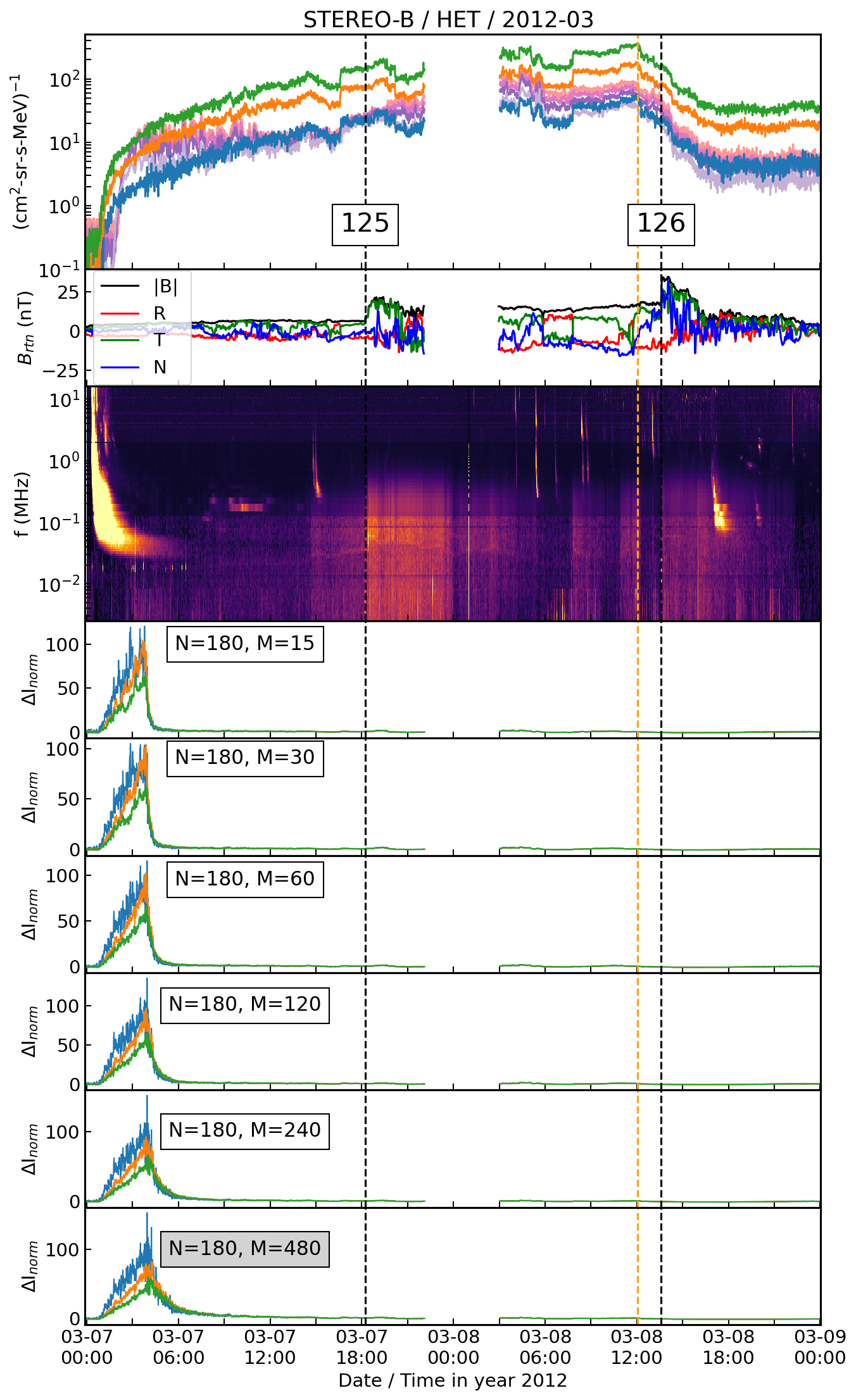}
\includegraphics[width=0.308\textwidth]{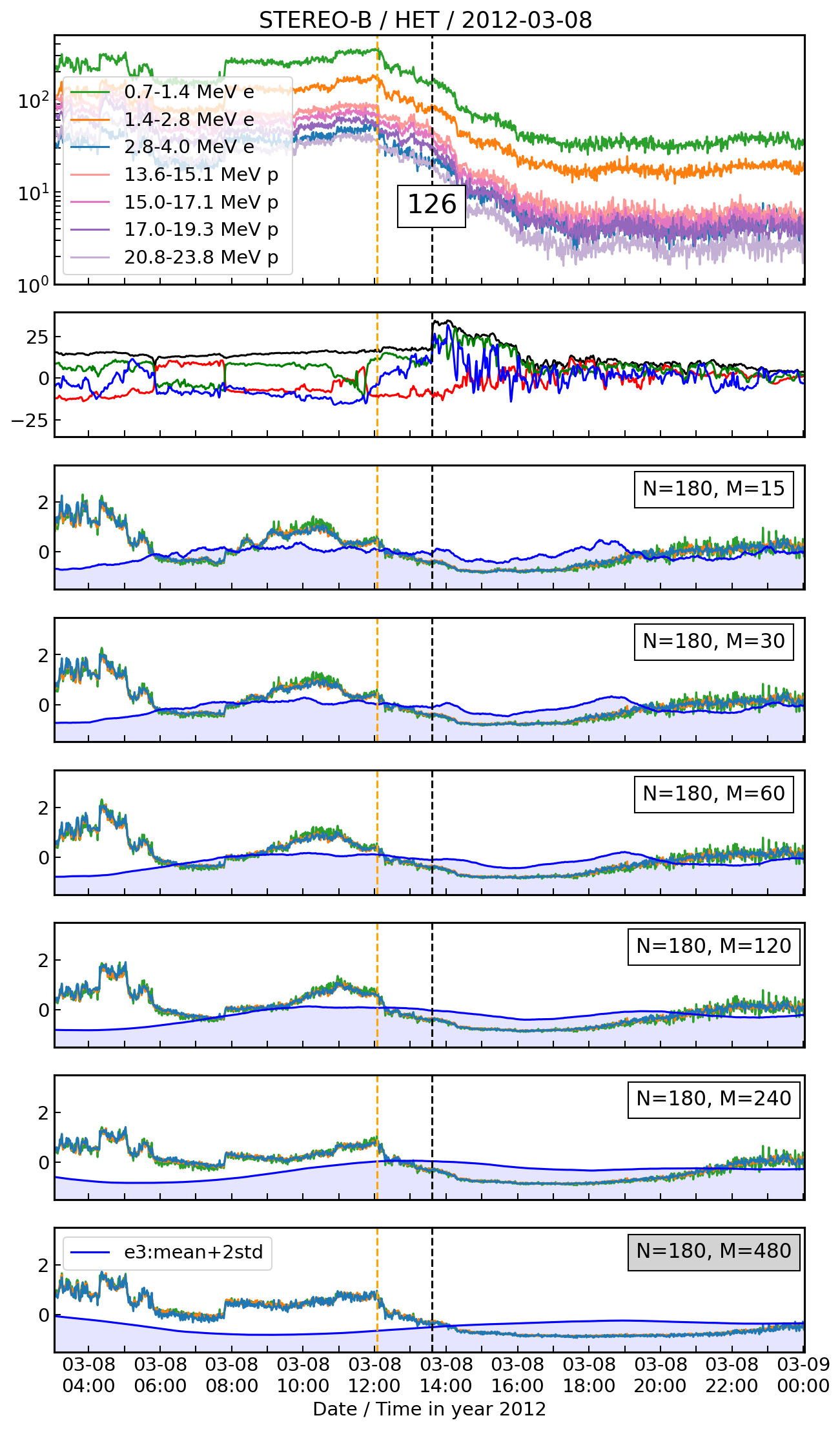} 
\includegraphics[width=0.30\textwidth]{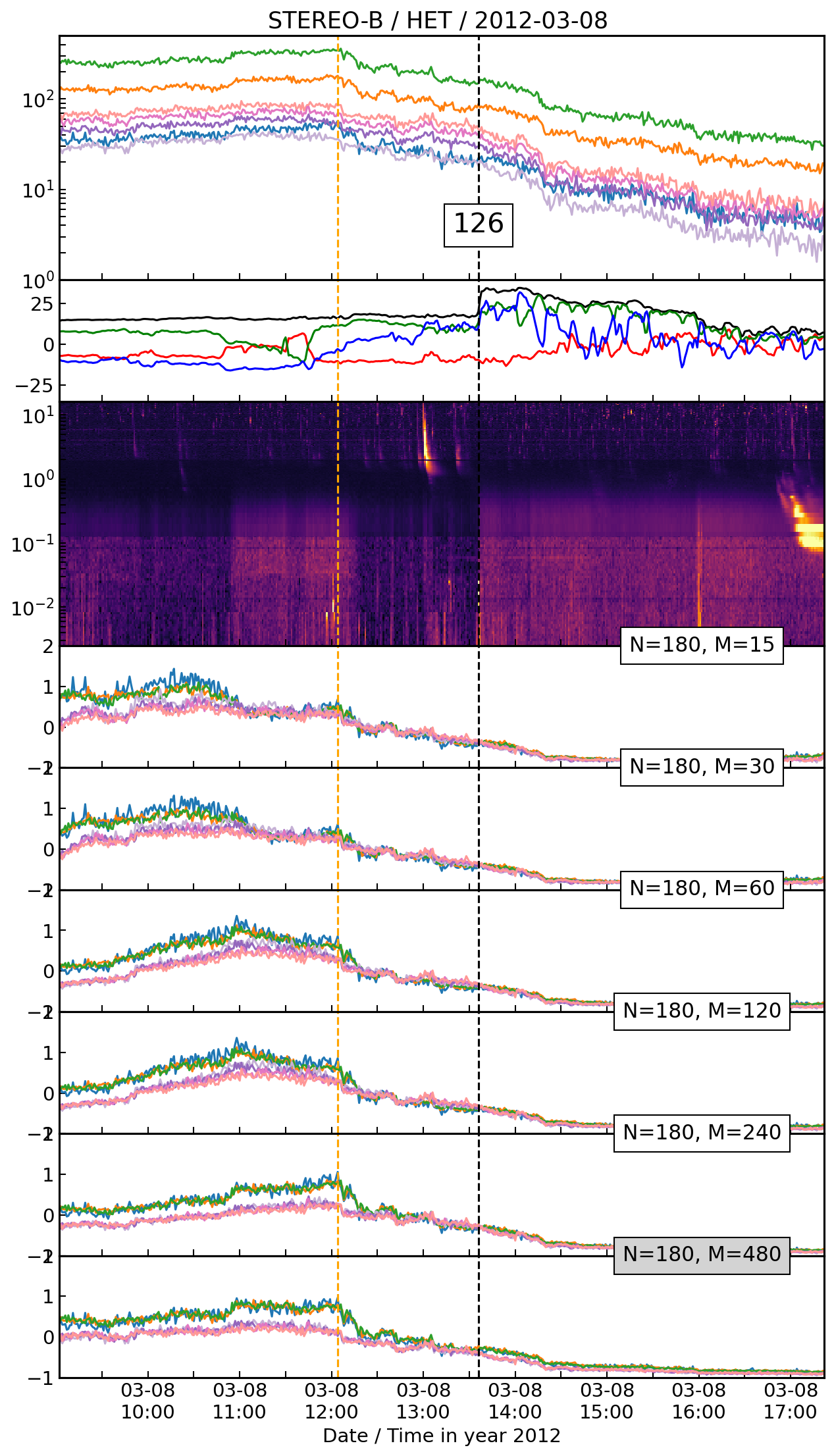}
\caption{Same format as Fig.~\ref{fig:141} but showing STEREO~B/HET observations of an SIR shock on 7 Mar 2012 and a ICME-driven shock on 8 Mar 2012.}
\label{fig:126}
\end{figure*}


\begin{figure*}
\centering
\includegraphics[width=0.31\textwidth]{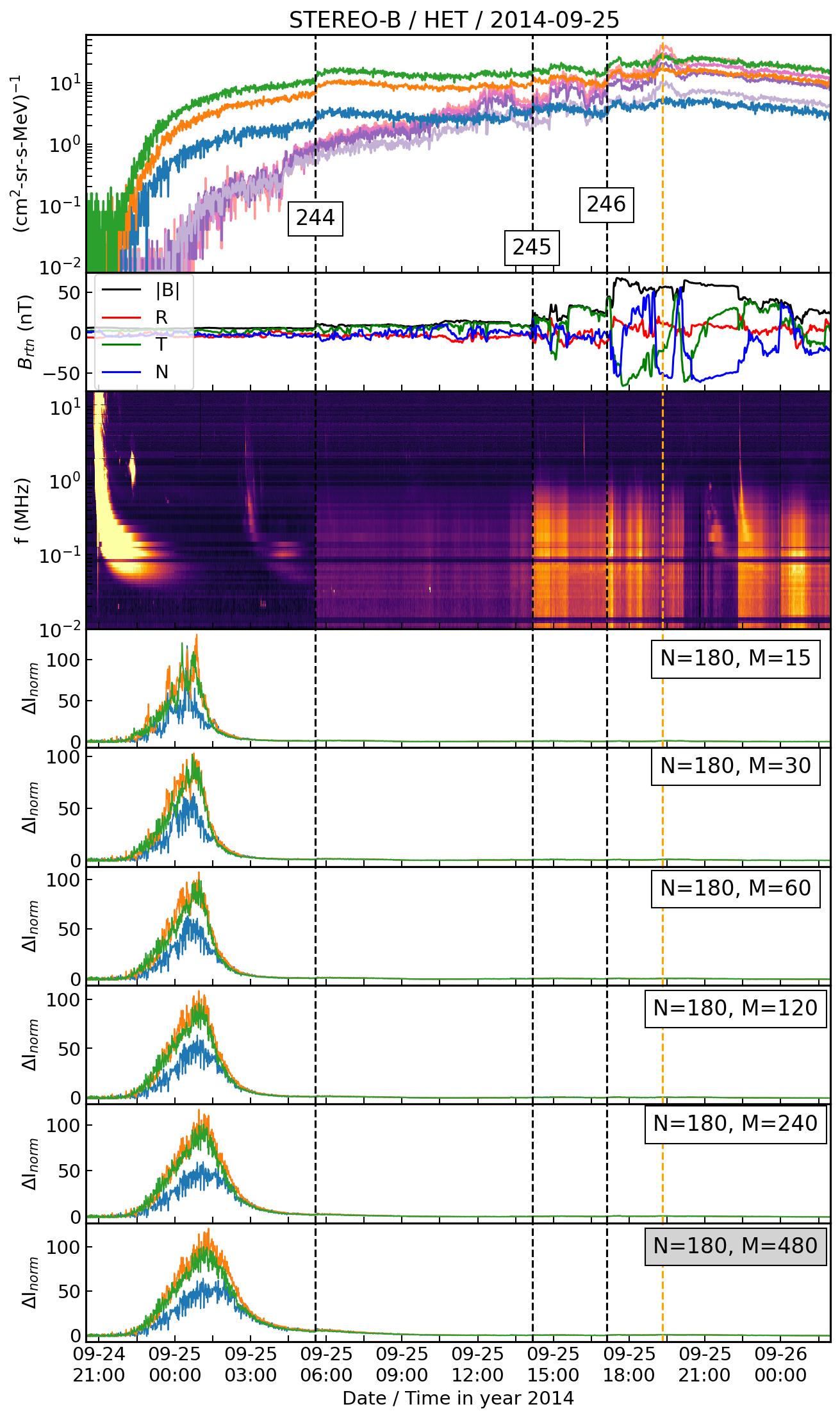}
\includegraphics[width=0.305\textwidth]{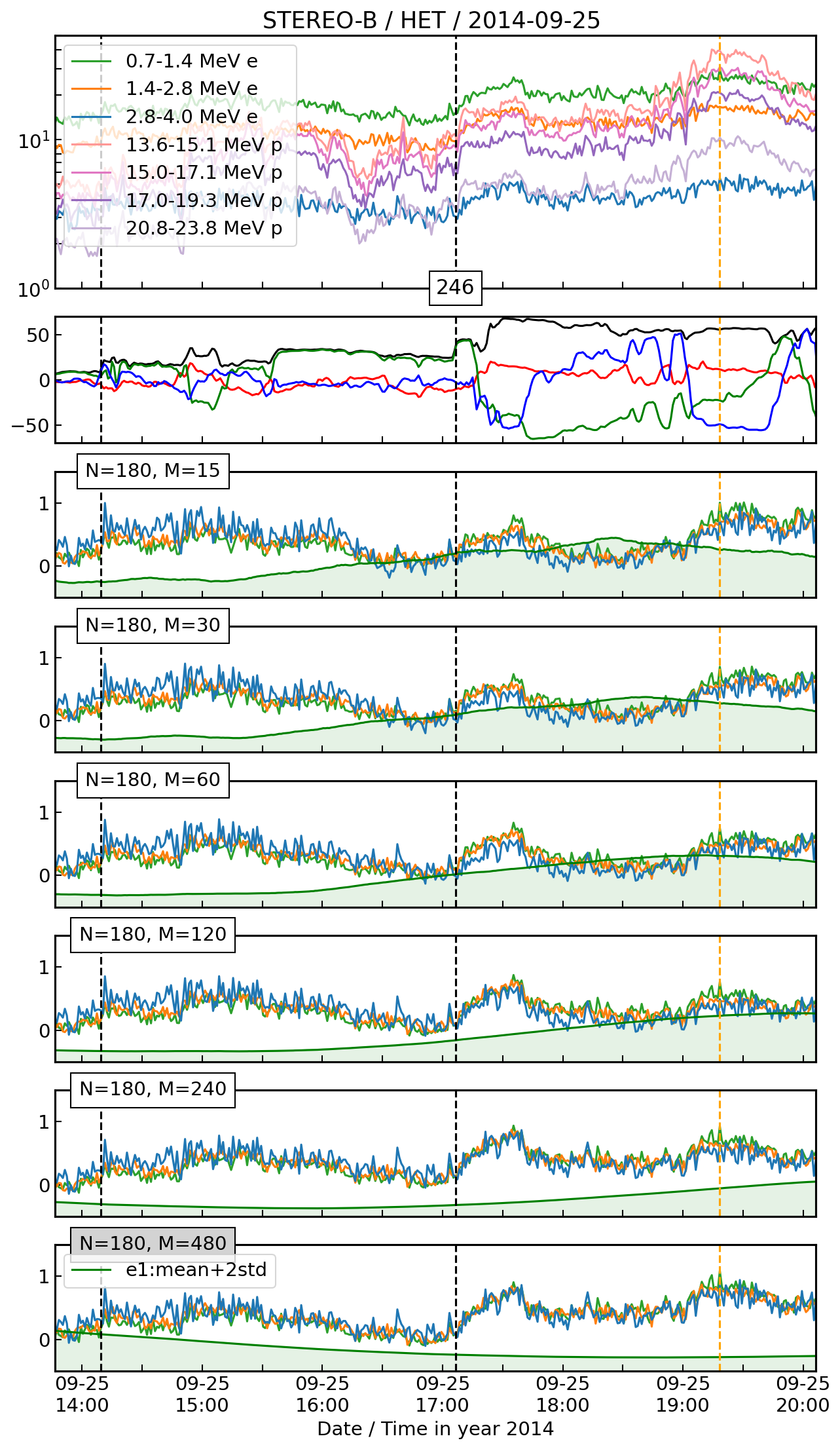} 
\includegraphics[width=0.308\textwidth]{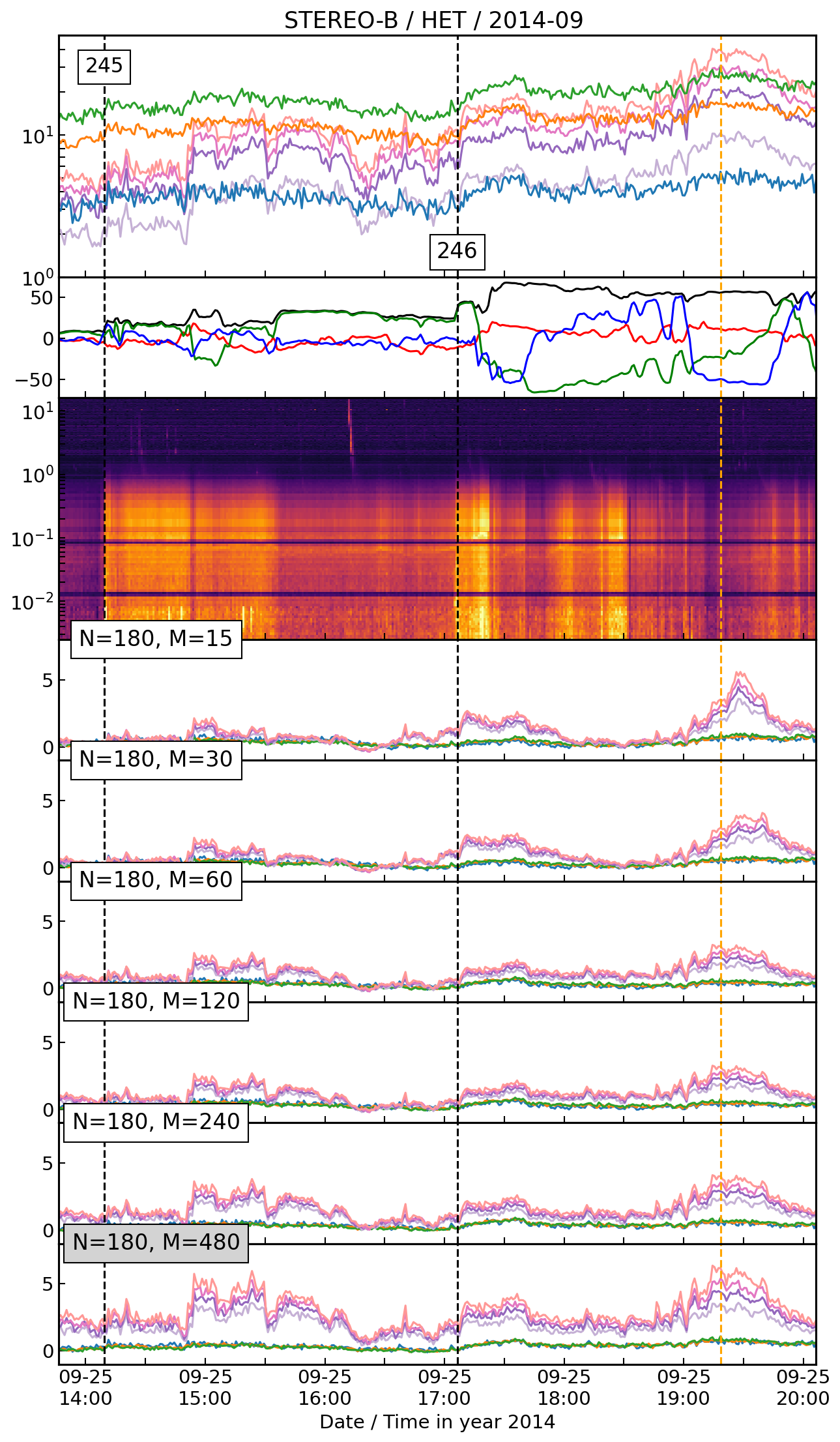}
\caption{Same format as Fig.~\ref{fig:141} but showing STEREO~B/HET observations of three ICME-driven shocks in one day on 25 Sep 2014.}
\label{fig:246}
\end{figure*}


\begin{figure*}
\centering
\includegraphics[width=0.325\textwidth]{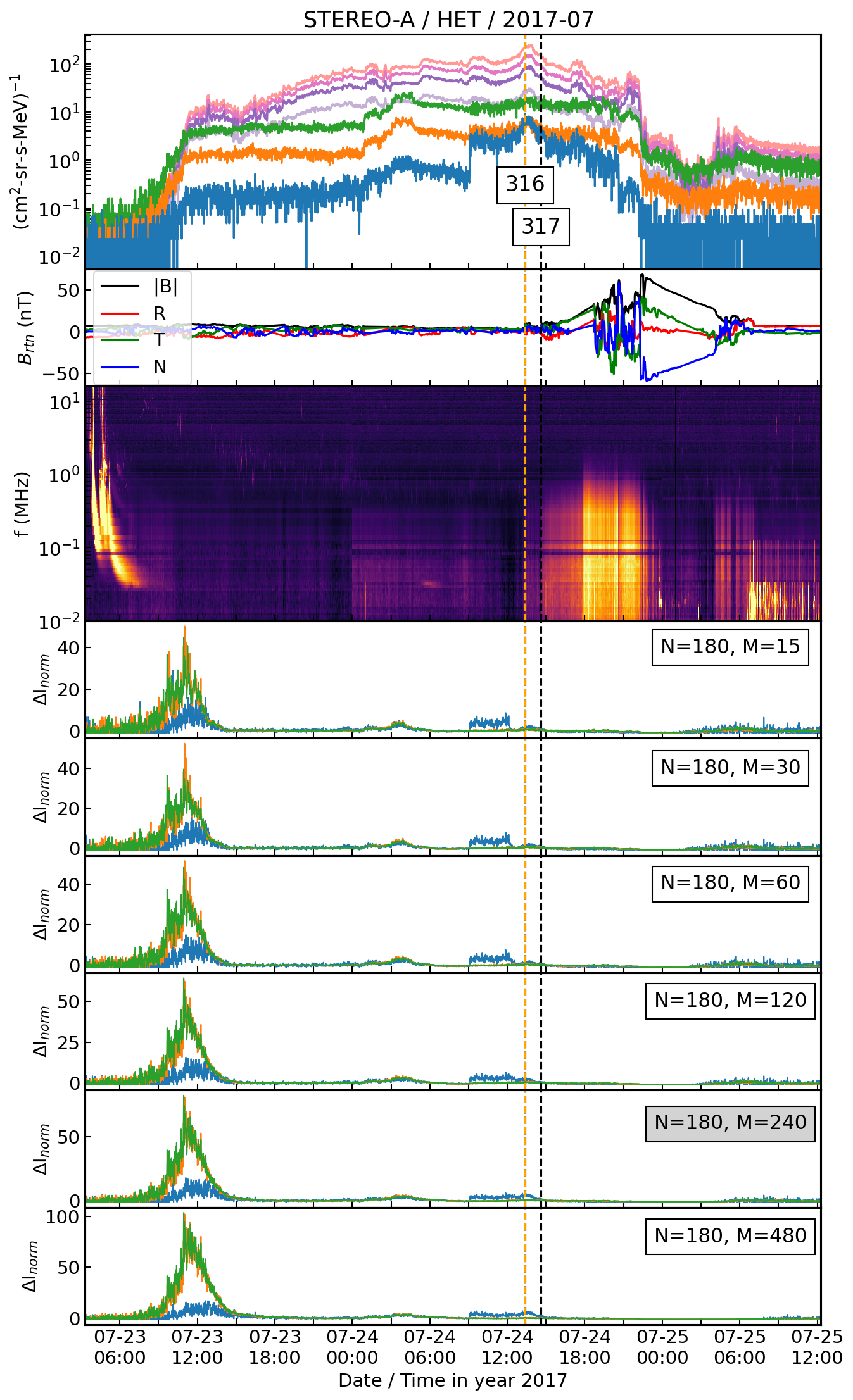}
\includegraphics[width=0.30\textwidth]{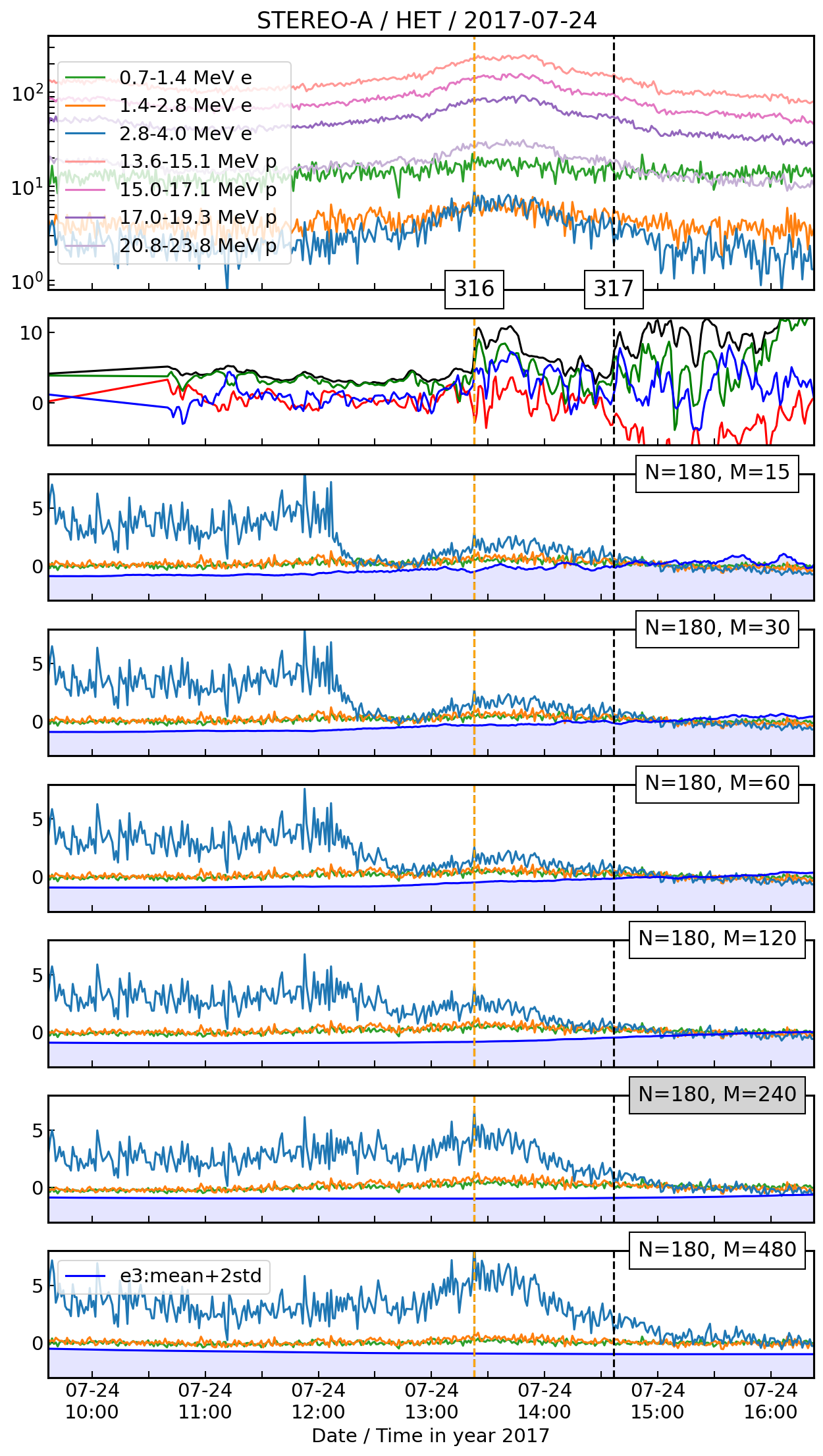} 
\includegraphics[width=0.304\textwidth]{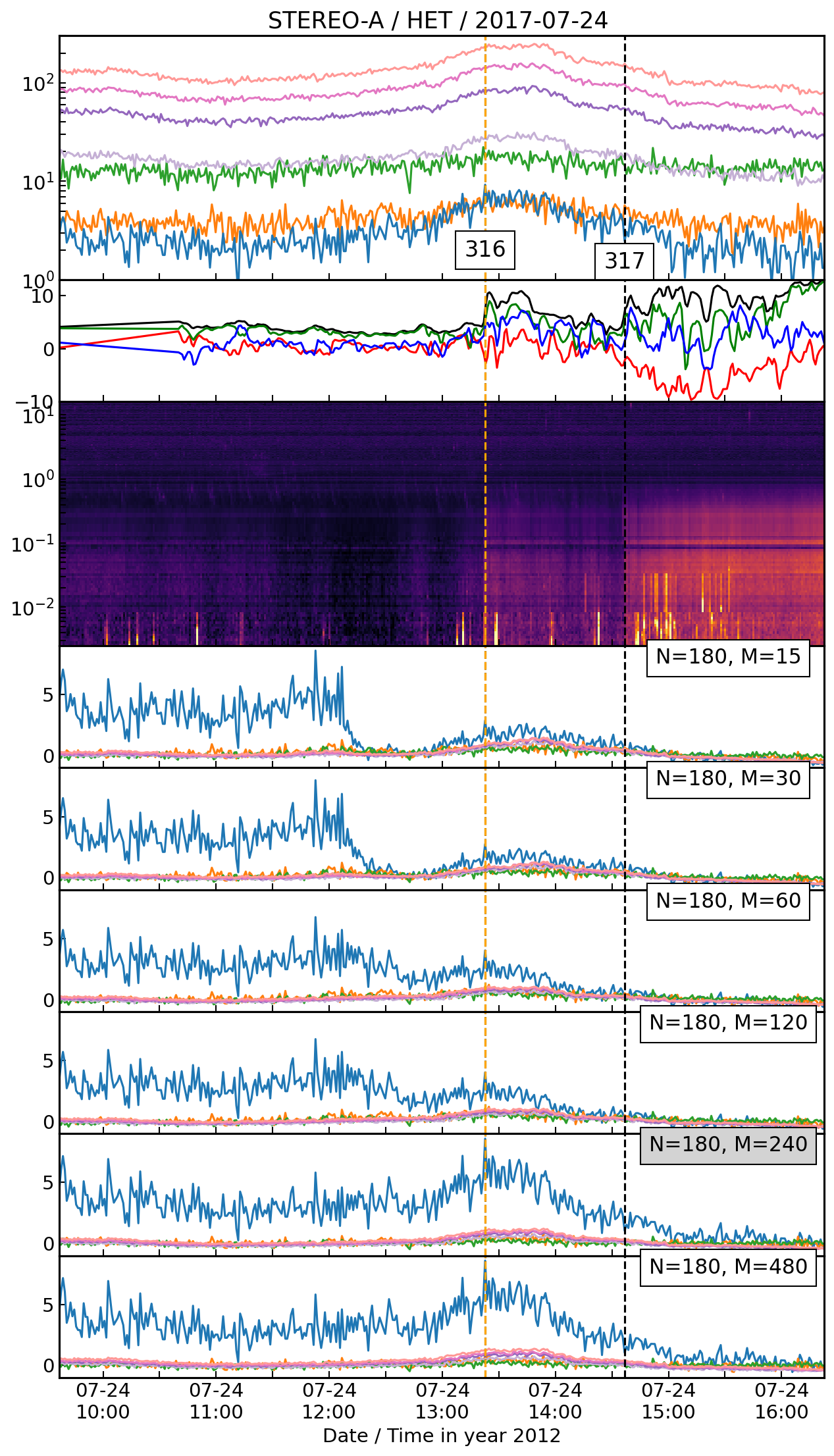}
\caption{Same format as Fig.~\ref{fig:141} but showing STEREO~A/HET observations of two ICME-driven shocks on 24 Sep 2017.}
\label{fig:316}
\end{figure*}
\begin{figure*}
\centering
\subfigure(a){\includegraphics[width=0.32\textwidth]{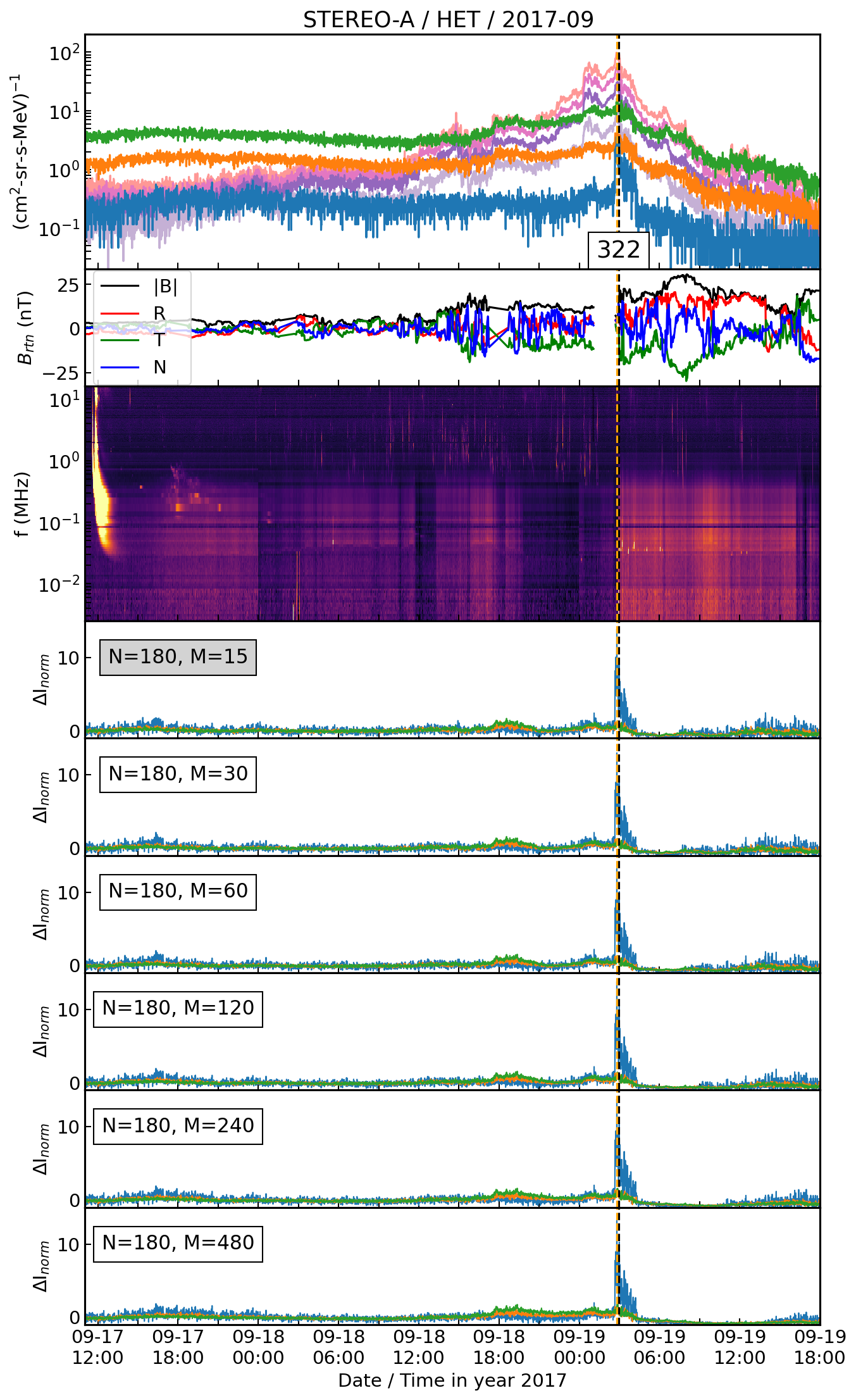}}
\subfigure(b){\includegraphics[width=0.30\textwidth]{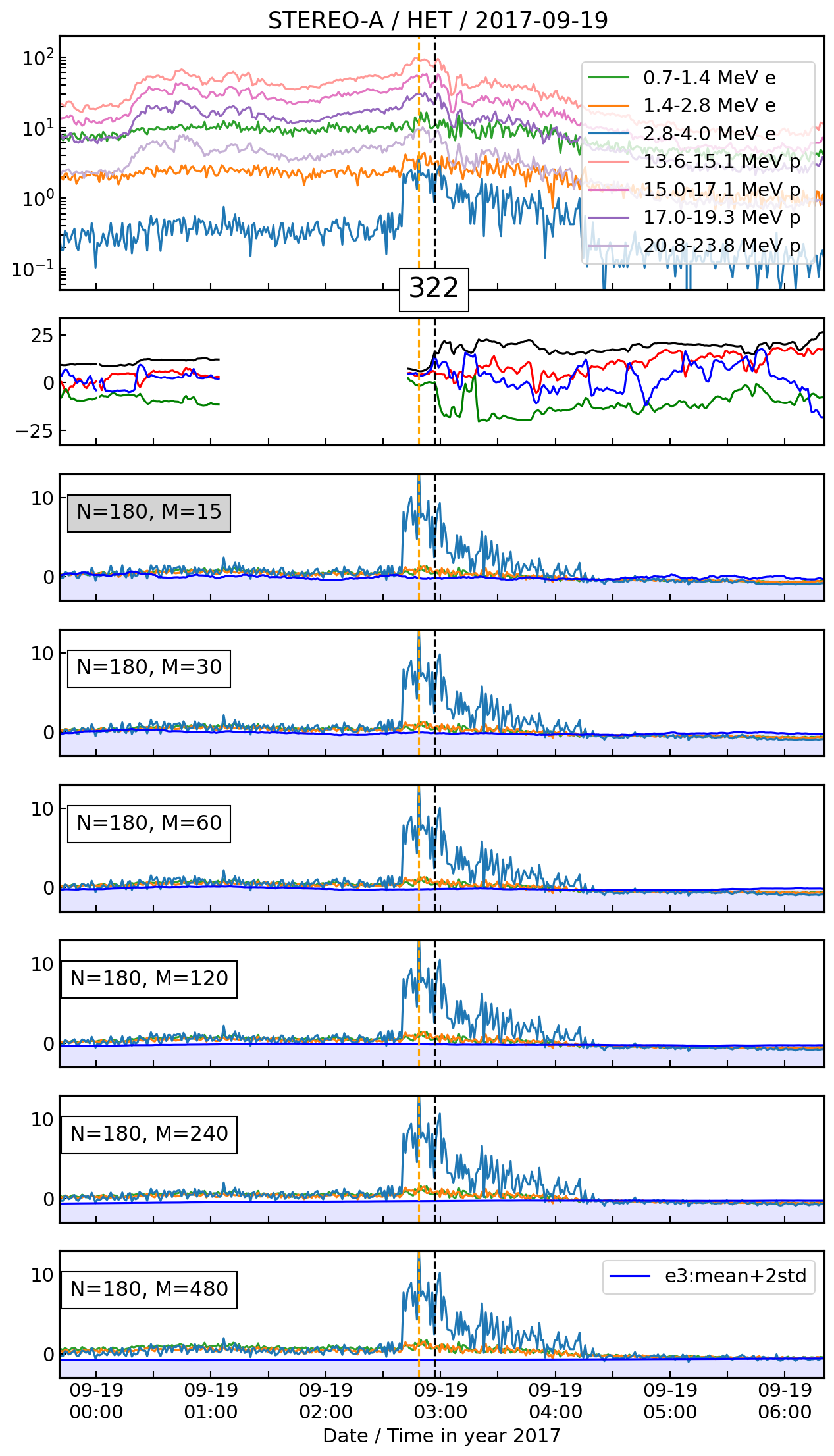}}
\subfigure(c){\includegraphics[width=0.30\textwidth]{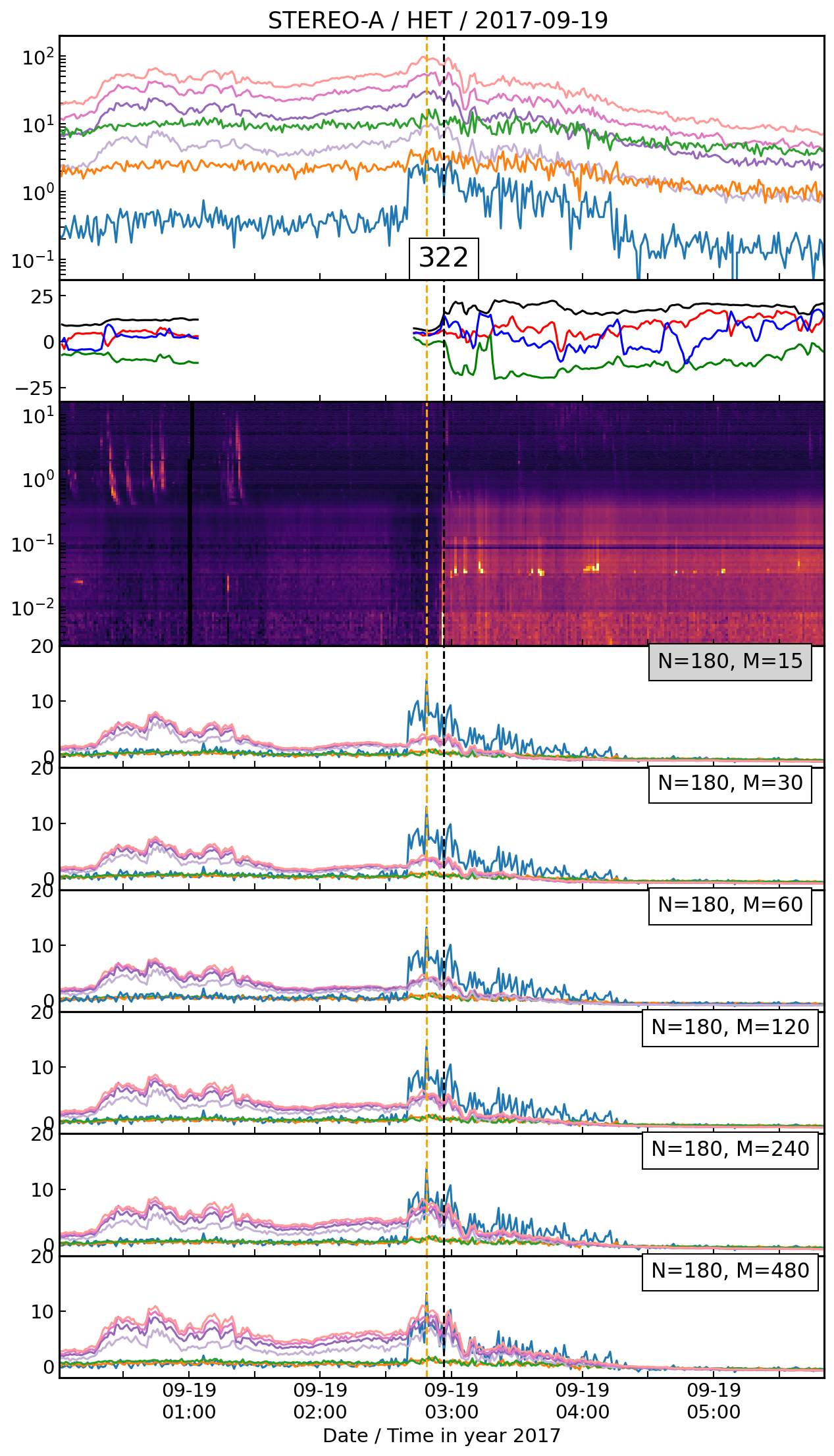}}
\caption{Same format as Fig.~\ref{fig:141} but showing STEREO~A/HET observations of an ICME-driven shock on 19 Sep 2017 at 2:56 with prominent plasma signatures on the downstream side of the shock.}
\label{fig:322}
\end{figure*}

\section{A test for proton contamination in the electron energy channels}\label{app:contamination}

\begin{figure*}
    \centering
    \includegraphics[width=0.33\textwidth]{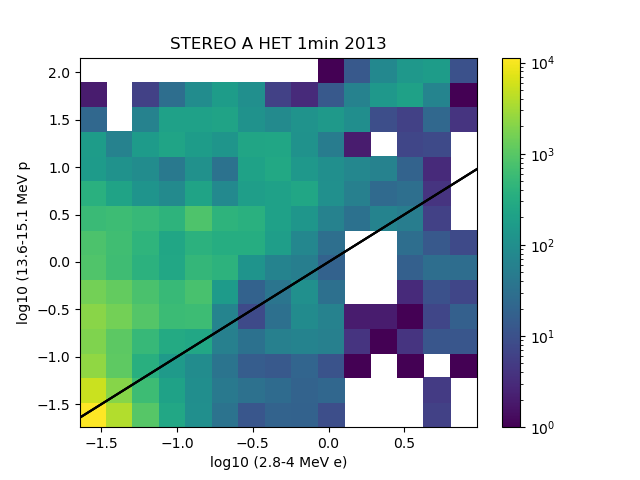}
    \includegraphics[width=0.33\textwidth]{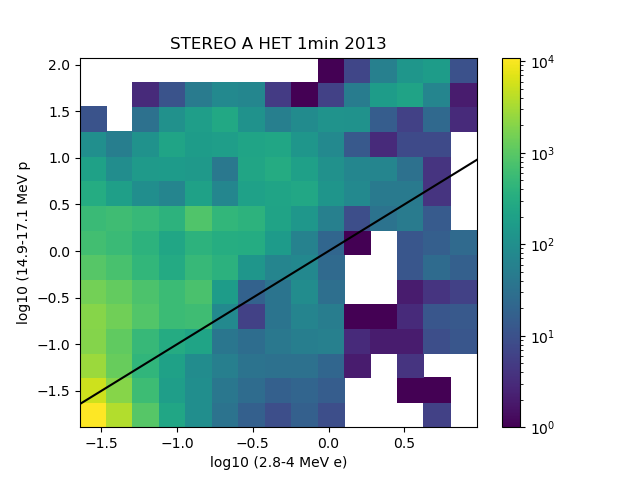}
    \includegraphics[width=0.33\textwidth]{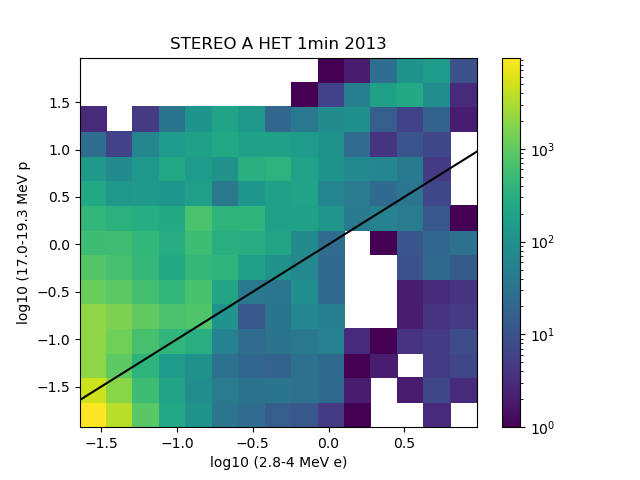}
    \caption{2D histograms of varying proton versus 2.8--4.0.~MeV electron intensities measured by STEREO-A/HET using 1-minute data of the whole year 2013. The plots show the logarithm of the intensities (excluding all zero intensities). The color scale (marking the number of observations in each bin) is logarithmic. The black solid line marks unity. The left, middle, and right panels show 13.6--15.1~MeV, 14.9-- 17.1~MeV, and 17.0--19.3~MeV protons, respectively.}
    \label{fig:Nina}
\end{figure*}

During periods of high proton intensities, such as in ESP events, the detection of small electron intensities can be difficult or even contaminated by the protons, which is a known feature in energetic particle instruments such as STEREO/SEPT \citep[e.g., ][]{Wraase2018}. To determine if there is proton contamination of the electron measurements used in this study we plot 2D histograms of deka-MeV proton intensities as a function of 2.8--4.0 MeV electron intensities in Fig \ref{fig:Nina}. We use 1-minute data of the whole year of 2013. None of the three proton channels shows a one-to-one correlation with the electron intensities. Furthermore, the low or even missing electron intensities at times when high proton intensities are observed (upper left corner), supports the assumption that proton contamination does not play a significant role in the electron channels.

\end{appendix}
\end{document}